\documentclass[%
preprint,
 amsmath,amssymb,
 aps,twocolumn,pre,notitlepage,10pt
]{revtex4-1}
\usepackage{graphicx} 
\usepackage{dcolumn} 
\usepackage{bm}
\usepackage{mathtools}
\usepackage{mhchem,xcolor}
\usepackage[paperwidth=210mm,paperheight=297mm,centering,hmargin=1.5cm,vmargin=2.5cm]{geometry}

\newcommand{\reactsh}[1]{\ensuremath{\xrightharpoonup{\hbox{\makebox[0.5cm][c]{\scriptsize $#1$}}}}}

\newcommand{\revreact}[2]{\ensuremath{\xrightleftharpoons[\hbox{\makebox[.25cm][c]{\scriptsize$#2$}}]{\hbox{\makebox[.3cm][c]{\scriptsize$#1$}}}}}

\begin{document}
\title{Complex Dynamics in Reaction-Phase Separation Systems}
\author{Dino Osmanovi\'c}
\affiliation
{Department of Mechanical and Aerospace Engineering, University of California at Los Angeles, Los Angeles, CA, USA}
\email{osmanovic.dino@gmail.com}
\author{Elisa Franco}
\affiliation
{Department of Mechanical and Aerospace Engineering, University of California at Los Angeles, Los Angeles, CA, USA}
\affiliation
{Department of Bioengineering, University of California at Los Angeles, Los Angeles, CA, USA}

%%%%%%%%%%%%%%%%%%%%%%%%%%%%%%%%%%%%%%%%%%%%%%%%%%%%%%%%%%%%%%%%%%%%

\begin{abstract}
 We investigate the emergence of sustained spatio-temporal  behaviors in  reaction-phase separation systems. We focus on binary systems, in which either one or both species can phase separate, and we discuss the stability of the homogeneous state determining the conditions for the emergence of a Hopf-type bifurcation. We then examine the effects of a specific autocatalytic chemical reaction, and computationally determine the full solutions to the partial differential equations. We find that when both species phase separate, sustained pulsed dynamics arise in one dimension. When considered in two dimensions, the system generates persistent, complex dynamic droplets, which do not generally appear if only one of the species can phase separate. We finally discuss the emergence of dynamics with complex features, which can be understood using the framework of a cellular automata.
\end{abstract}

\maketitle

\section{Introduction}

The possibility to couple phase separation with chemical reactions has commanded recent attention for its potential applicability toward controlling the intracellular environment through membraneless organelles~\cite{rai2018kinase,garabedian2021designer}, as well as the behavior of biomolecular materials through the emergence of dense compartments~\cite{agarwal2024dynamic,deng2020programmable}. From a theoretical perspective, reaction-phase separation systems are an interesting extension of reaction-diffusion systems, which are known to have the capacity to exhibit a variety of complex spatio-temporal patterns~\cite{ringkvist_dynamical_2009,zheng_pattern_2015,mcgough_pattern_2004}. 

Phase separated systems are amenable to being studied in terms of the existence of spatially extended patterns~\cite{kirschbaum2021controlling,Osmanovic2023,luo2023influence} as well as well the emergence of interface properties~\cite{bauermann_formation_2023,cho_nonequilibrium_2023}, however the question of dynamics in such systems is more involved. ``Exotic'' behaviors have been observed, such as droplet splitting~\cite{zwicker_2016} and droplet appearance and disappearance~\cite{Li2020}, but few studies have considered the combination of aspects which we deem \textit{uniqueness}, \textit{complexity}, and \textit{persistence}.

By uniqueness, we mean to what extent a dynamic behavior may arise when species phase separate, but not when they are merely diffusive. Such a question is generally challenging to address, as the additional nonlinearities and higher order terms introduced by phase separation make analytical treatments more difficult.  For example, droplet splitting has been observed in the FitzHugh-Nagumo equations~\cite{pearson_complex_1993,lee_lamellar_1995}, hence splitting need not necessarily be unique to phase separating systems in general. However, in the presence of specific reactions, certain dynamics may be exclusive to phase separation and cease to occur in its absence. 

Complex dynamics are characterized by broad ranges of length and timescales. What complex dynamic behaviors can appear in reaction-phase separation systems? For example, a behavior of particular interest is the spontaneous appearance and disappearance of dense droplets under the influence of chemical reactions. Indeed, phase separating systems have been implicated as potential model systems for "life-like" behaviors~\cite{demarchi_enzyme-enriched_2023,zwicker_droplets_2023}, where temporal fluctuations of condensates is key.

%What complex dynamic behaviors can appear in reaction-phase separation systems? A behavior of particular interest is the spontaneous appearance and disappearance of dense droplets under the influence of chemical reactions, which can be viewed as an elementary step toward the emergence of cycles of life. Indeed, phase separating systems have been implicated as potential model systems for "life-like" behaviors~\cite{demarchi_enzyme-enriched_2023,zwicker_droplets_2023}, where temporal fluctuations of condensates is key.  

Finally, can complex dynamics that are unique to reaction-phase separation  also be persistent? This question is relevant because a complex dynamics  could be observed transiently but disappear over long timescales. Alternatively, the dynamics could only appear under special choices of the initial conditions or boundary conditions. In other words, it is significant to investigate what are the possible types of \textit{sustained} dynamics that can appear in these systems, and could potentially give rise to long term, autonomous behaviors reminiscent of replication and evolution~\cite{zwicker_droplets_2023}. If we think of living behaviors as persistent, complex dynamics of droplets, then such investigations may be related to the emergence of spatiotemporal chaos in reaction-diffusion systems\cite{rossler_chemical_1976,zimmermann_pulse_1997,strain_size-dependent_1998,cai_spatiotemporal_2001,castelino_spatiotemporal_2020}.
Perhaps even more speculatively, discrete models such as cellular automata can also exhibit life-like behaviors and emergence through simple deterministic rules~\cite{sayama_self-reproduction_2024}, and the possibility that interacting phase separated droplets could lead to similar emergence is an intriguing one.

In this paper, we investigate which dynamic behaviors can emerge in a reaction-phase separation system by focusing our attention on a two-species system that are coupled via chemical reactions for a random field initial condition. We consider the cases in which either both or only one of the species can phase separate (while the other is simply a diffusive species).  We then review a general recipe to  investigate the emergence of sustained oscillatory dynamics in this two-species system, using a linearized approach. Finally, we computationally explore the behavior of the partial differential equations associated with our models, to determine the actual forms of dynamics that arise in these systems. We find that complex dynamics behaviors, including sustained birth-death events and spatial fluctuations, are more likely to arise when both species are phase separators and there is an energetic penalty to their co-localization. These results suggest that competition for space may be a key ingredient for the emergence of complex behaviors in living matter.

\section{Model reaction-phase separation systems and variants}

We introduce the model of a reaction-phase separation system with two different variants. Each variant includes two components, or chemical species,  $P$ and  $Z$.  In model (M1), we allow species $P$ to phase separate, and designate species $Z$ as an exclusively diffusive species obeying standard reaction-diffusion equations. In model (M2) we allow both components $P$ and $Z$ to phase separate independently.

The free energy describing phase separation of a molecular component is given by the Landau-Ginzburg functional:
\begin{equation} \label{eq:LGfunc}
F_{\text{ps}}[c]=\int \mathrm d \mathbf x \nu(c(\mathbf x,t)-\rho_{0})^2(c(\mathbf x,t)-\rho_{1})^2+\gamma^2 |\nabla c_(\mathbf x,t)|^2,    
\end{equation}
where $c(x,t)$ is the concentration field of the component, $\mathbf x$ is a spatial coordinate vector, $\nu$ is some energy scale, $\rho_{0}$ and $\rho_{1}$ are the respectively the disperse and dense phase concentrations, and $\gamma$ is a surface tension parameter.

The behavior of a purely diffusive component, that does not undergo phase separation, is described by a simple free energy expression: 
\begin{equation} \label{eq:LGdiff}
F_{\text{d}}[c]=\xi \int \mathrm d \mathbf x c(\mathbf x,t)^2.   
\end{equation}

Because in our system we consider two components, $P$ and $Z$, we also allow them to have conserved interactions beyond chemical reactions. We model these interactions through the following free energy: 
\begin{equation}
    F_{\text{int}}[c_P,c_Z]=\chi \int \mathrm d \mathbf x c_P(\mathbf x,t)c_Z(\mathbf x,t),
\end{equation}
where $\chi$ is an energy scale that gives the magnitude of the free energy penalty to overlapping concentrations of $c_P$ and $c_Z$. Therefore, when $\chi>0$ colocalization of $P$ and $Z$ is disfavored.

The two free energy models are:
\begin{equation}\label{eq:M1} \tag{M1}
    F_{M1}=F_{\text{ps}}[c_P]+F_{\text{d}}[c_Z]+F_{\text{int}}[c_P,c_Z], 
\end{equation}
\begin{equation} \label{eq:M2} \tag{M2}
    F_{M2}=F_{\text{ps}}[c_P]+F_{\text{ps}}[c_Z]+F_{\text{int}}[c_P,c_Z]. 
\end{equation}
% $F^{(i)}=F_{ps}[c_A]+F_{ps}[c_B]+F_{\text{int}}[c_A,c_B]$ and $F^{(ii)}=F_{ps}[c_A]+F_D[c_B]+F_{\text{int}}[c_A,c_B]$. 

The time evolution of components $P$ and $Z$ can be derived via the so-called model B  dynamics\cite{Hohenberg1977}:
\begin{align} \label{eq:fund}
    &\frac{\partial c_P(\mathbf x,t)}{\partial t}=D_P\nabla^2 \left(\frac{\delta F_{Mi}(c_P,c_Z)}{\delta c_P}\right)+R_P(c_P,c_Z), \\
    &\frac{\partial c_Z(\mathbf x,t)}{\partial t}=D_Z\nabla^2 \left(\frac{\delta F_{Mi}(c_P,c_Z)}{\delta c_Z}\right)+R_Z(c_P,c_Z),
\end{align}
where $F_{Mi}$ is either the free energy $F_{M1}$ or $F_{M2}$, and where we introduced chemical reaction terms $R_P,R_Z$ that describe production, degradation, or inter-conversion of $P$ and $Z$ (according to the law of mass action). We make two key assumptions constraining the types of chemical reactions we consider: 1) we assume there exists a non-zero stable fixed point $(c_P^*, c_Z^*)$ where $R_P(c_P^*,c_Z^*)=R_Z(c_P^*,c_Z^*)=0$, and 2) that there is mass conservation in the dynamics described by $R_P$ and $R_Z$. Finally, we also assume that the mobility matrix is diagonal, such that we obtain the two coefficients $D_P$ and $D_Z$. 

To facilitate further analysis, we non-dimensionalize the two models, as shown in the Supplementary Information (SI) Section I, and we obtain a set of two nondimensional equations:
{\small \begin{align} \label{eq:fund1}
    &\frac{\partial \tilde c_P}{\partial \tau}=\nabla^2 \left(\Pi_P(\tilde c_P)+(\chi/\rho) \tilde c_Z\right)-\nabla^4 \tilde c_P + \tilde R_P(\tilde c_P,\tilde c_Z), \\  \label{eq:fund2}
    &\frac{\partial \tilde c_Z}{\partial \tau}=\delta \nabla^2 \left(\Pi_Z(\tilde c_Z)+\chi \rho \tilde c_P(\mathbf x,t)\right)-\delta g^2 \nabla^4 \tilde c_Z+\tilde R_Z(\tilde c_P,\tilde c_Z),
\end{align}}
where a tilde represents a nondimensional variable, and $\Pi$ is a nonlinear function that represents the propensity of the system to phase separate.  

We highlight  two important parameters that we will examine in detail later:
\begin{align}
    \delta &:= D_Z/D_P, \\
    g &:= \gamma_Z/\gamma_P, 
\end{align}
 which respectively capture the ratio of diffusivities ($\delta$) and the ratio of surface tensions ($g$) of  species $P$ and $Z$.  Equations~\eqref{eq:fund1}-\eqref{eq:fund2} specialize in model (M1), where only species $P$ phase separates, by taking $g\to0$ and $\Pi_z(\tilde c_Z)\to \tilde c_Z$.  Otherwise, we obtain model (M2) in which both $P$ and $Z$ phase separate. 

%Equations \eqref{eq:fund} and \eqref{eq:fund2} can be solved numerically.

\section{Linear Stability Analysis }
Linearization of the model~\eqref{eq:fund1}-\eqref{eq:fund2} is the  simplest approach to investigate the appearance of dynamics. We perturb the homogeneous solution near the chemical equilibrium, substituting $(\tilde c_P(\mathbf x, t),\tilde c_Z(\mathbf x, t))=(\tilde c_P^*,\tilde c_Z^*)+\boldsymbol \epsilon \exp(i \mathbf k.\mathbf x +\beta t)$ into Equations~\eqref{eq:fund1}-\eqref{eq:fund2} ($\boldsymbol \epsilon$ is a small vector, $\mathcal{O}(\boldsymbol \epsilon^2)\sim0$), and we obtain a relationship between the growth rate $\beta$ and the wavelength $\mathbf k$:
\begin{align} \label{eq:eigenvalue}
\beta\boldsymbol \epsilon &= 
\begin{bmatrix}
 a_{\text{PP}}(\mathbf k) &  a_{\text{PZ}}(\mathbf k) \\
\delta \, a_{\text{ZP}}(\mathbf k) & \delta \, a_{\text{ZZ}}(\mathbf k) \\
\end{bmatrix}.\boldsymbol{\epsilon}
+\begin{bmatrix}
J_{\text{PP}} & J_{\text{PZ}} \\
J_{\text{ZP}} & J_{\text{ZZ}} \\
\end{bmatrix}.\boldsymbol\epsilon,
\\ \nonumber
&= (\underline \Delta(\mathbf k) + \underline J ).\epsilon=\underline M (\mathbf k).\epsilon,
\end{align}
where the first matrix $\underline \Delta(\mathbf k)$ arises from the phase separation terms operated on by $\nabla^2,\nabla^4$ and the second matrix $\underline J$ arises from the reaction terms $\tilde R_P, \tilde R_Z$. The individual entries of these matrices are: 
\begin{align}
a_{\text{PP}}(\mathbf k)=& -\mathbf k^2\left( \frac{\mathrm d \Pi_P(\tilde c_P)}{\mathrm d \tilde c_P}\bigg|_{\tilde c_P*}\!\!\!+k^2\right) \\  \nonumber
 \coloneqq &-\mathbf k^2\left(p_a+\mathbf k^2\right),\\
a_{\text{PZ}}(\mathbf k)=&-(\chi/\rho) \mathbf k^2,\\
a_{\text{ZP}}(\mathbf k)=&-(\chi \rho) \mathbf k^2,  \label{eq:offdiagonal} \\
a_{\text{ZZ}}(\mathbf k)=& -\mathbf k^2\left( \frac{\mathrm d \Pi_Z(\tilde c_Z)}{\mathrm d \tilde c_Z}\bigg|_{\tilde c_Z*}+ g^2 \mathbf k^2\right)\\ \nonumber 
\coloneqq & -\mathbf k^2\left(p_b+g^2 \mathbf k^2\right), \\
J_{ij} =&\frac{\partial \tilde R_i(\tilde c_P,\tilde c_Z)}{\partial \tilde c_j}\bigg|_{{\tilde c_P^*},{\tilde c_Z^*}}, \, i=P,Z,\, j=Z,P. 
\end{align}
We now focus our attention on the emergence of oscillations near equilibrium, which may predict sustained dynamics in $\tilde c_P$ and $\tilde c_Z$. We can evaluate whether temporal oscillations occur by examining the eigenvalue Equation~\eqref{eq:eigenvalue}. Oscillations near equilibrium occur when the dominant eigenvalue $\lambda_{\text{max}}$ satisfies \cite{Cross1993}:
\begin{equation}\label{eq:cond}
\text{Re}(\lambda_{\text{max}})>0,\, \text{Im}(\lambda_{\text{max}})\ne0.
\end{equation} 
Because this system only includes two variables, we can derive exact expressions for the eigenvalues corresponding to Equation~\eqref{eq:eigenvalue}:  
\begin{align}
&\lambda=\frac{1}{2}\left(\text{Tr}(\underline M(\mathbf k) )\pm\sqrt{\text{Tr}(\underline M)^2-4\text{det}(\underline M(\mathbf k))}\right),
\end{align}
therefore condition~\eqref{eq:cond} is equivalent to the inequalities: 
\begin{align} \label{eq:trace}
&\text{Tr}(\underline M(\mathbf k)) >0, \\ \label{eq:det}
&\text{det}(\underline M(\mathbf k)) >\frac{1}{4}\text{Tr}(\underline M(\mathbf k))^2,
\end{align}
where we can look for $\mathbf k$ that satisfies them. In the next section, we discuss the influence of specific chemical reactions involving $P$ and $Z$ on Equations~\eqref{eq:trace} and~\eqref{eq:det}. The parameters associated with chemical reactions can lead to the emergence of sustained oscillations near equilibrium if condition~\eqref{eq:eigenvalue} is satisfied, introducing a Hopf-type bifurcation~\cite{marsden2012hopf}.   

\section{Candidate Chemical Reactions for Sustained Dynamics}

The stability analysis in the previous section provides conditions and assumptions that are useful to screen and identify candidate chemical reactions that, coupled with phase separation, may yield oscillatory dynamics. The conditions are listed below: 
%\boldsymbol{v^T} .\underline J =
\begin{align}
\text{det}(\underline J)&=0 \text{ (mass conservation) } \label{eq:cond1}\\
\text{Tr}(\underline J)&<0 \,\text{ (stable chemical fixed point)} \label{eq:cond2}\\
\text{Tr}(\underline \Delta(\mathbf k) + \underline J) &>0 \text{ (growth instability) } \label{eq:cond3}\\
\text{det}(\underline \Delta(\mathbf k) + \underline J) &>\frac{1}{4}\text{Tr}(\underline \Delta(\mathbf k) + \underline J)^2 \text{ (oscillations) } \label{eq:cond4}
\end{align}
 The combination of conditions~\eqref{eq:cond2} and~\eqref{eq:cond3} suggests we must be in a regime where phase separation would occur in the absence of chemical reactions. 
%(Due to the fact that all entries of  matrices $\underline \Delta(\mathbf k)$ and $\underline J$ are real, condition~\eqref{eq:cond4} also tells us that $\text{det}(\underline \Delta(\mathbf k) + \underline J)$  must be positive.) 
All these conditions can be analyzed in terms of individual 
parameters that describe phase separation and chemical reactions. These algebraic expressions are lengthy, and we chose to interrogate them with a graphical approach described in detail in SI Section III, following our previous work~\cite{Osmanovic2023}. 

The most salient features emerging from our analysis are that a Hopf-type bifurcation is more likely to occur when $\chi>0$ (mixing of $P$ and $Z$ is disfavored) and when $J_{\text{PP}}$ and $J_{\text{ZZ}}$ have different signs or different magnitudes SI Fig.s 1-3. This scenario is unlikely in our system (closed with mass conservation), because  $J_{\text{PP}}$, $J_{\text{ZZ}}$ are less than or equal to zero near the chemical equilibrium. 

We can obtain simpler expressions from the outset by assuming condition \eqref{eq:cond4} is true. In this case, a Hopf-type bifurcation occurs in this system when:
\begin{equation}
J_{\text{PP}}+J_{\text{PZ}}+\frac{(p_a+\delta p_b)^2}{4(1+g^2 \delta)}>0,
\end{equation}
at a wavelength $k=\frac{\sqrt{-p_a-\delta  p_b}}{\sqrt{2} \sqrt{\delta  g^2+1}}$.

Finally, we note that expression \eqref{eq:cond4}, while difficult to analyze for all parameters, can be more easily interpreted in terms of what choice of $\chi$ is favorable to dynamics. We recall that $\chi$ is an energy scale proportional to the free energy associated to colocalization of $P$ and $Z$. For example, for a $2\times 2$ real matrix $\underline M$=
$\begin{bmatrix}
    m_1 & m_2 \\ m_3 & m_4
\end{bmatrix}$, condition~\eqref{eq:cond4} is equivalent to $4 m_2 m_3+(m_1-m_4)^2<0$. Term $(m_1-m_4)^2$ is positive because the matrix is real, therefore $m_2 m_3$ must be negative to satisfy the condition. In our system this means that:
\begin{equation}
    -\delta  k^2 \chi  J_{\text{PZ}}-k^2 \chi  J_{\text{ZP}}+J_{\text{PZ}}
   J_{\text{ZP}}+\delta  k^4 \chi ^2 <0.
\end{equation}
Here $\delta>0$, and for a closed system we have that $J_{\text{PZ}}\le0,J_{\text{ZP}}\le0$. For these reasons, $\chi$ must be positive for the emergence of a Hopf-type bifurcation, i.e., the mixing of of $P$ and $Z$ should be energetically penalized. 

In the next section, we will examine a specific network that locally satisfies conditions~\eqref{eq:cond2} and~\eqref{eq:cond4}, and examine whether oscillatory behaviors emerge from the numerical integration of the full nonlinear model.

\section{Computational Simulation Results}

\subsection{A chemical network generating oscillatory phase separation patterns} 
Thus far, we kept our discussion general with a consideration of linearized forms of chemical dynamics. Here we narrow our focus on a particular kind of chemical reaction and we test the kinds of dynamics that can occur in such a system. We consider the two species network:
\begin{align}\label{eq:C1}
\ce{P + Z & \reactsh{k_1} Z + Z}, \tag{C1}\\
\ce{Z &\reactsh{k_2} P}. \nonumber
\end{align}

There are several reasons we chose these particular chemical reactions. While theoretically, we can tune our system parameters to whichever parameter region has the oscillatory instability, in real systems there is often less latitude to modify individual parameters (notwithstanding the fact that these equations are already a large simplification of real systems). As compared to linear reactions (i.e., $A\revreact{}{} B$), the chemical reactions~\eqref{eq:C1} have the feature that the elements of the Jacobian automatically satisfy the aforementioned condition that they are of different magnitudes, as $J_{\text{ZZ}}=0$ around the fixed point of the chemical dynamics.

Moreover, the extensive literature on pattern formation in reaction diffusion systems often introduce chemical reactions that would introduce some nonlinearity and positive feedback loops into the underlying equations~\cite{gierer_theory_1972,landge_pattern_2020,seshasai_multiplicity_2020}. While phase separating systems are already nonlinear, we anticipate that nonlinear chemical dynamics would yield more interesting dynamical possibilities than linear chemical reactions. The chemical reactions chosen bear some resemblance to the Gray-Scott model~\cite{mcgough_pattern_2004,gray_autocatalytic_1983}, and is a simple example of an auto-catalytic system. The reactions in~\eqref{eq:C1} also present a correspondence to the susceptible-infected-susceptible (SIS) model of epidemic spread~\cite{hethcote_three_1989}. 

\begin{figure}[htbp]
    \includegraphics[width=\columnwidth]{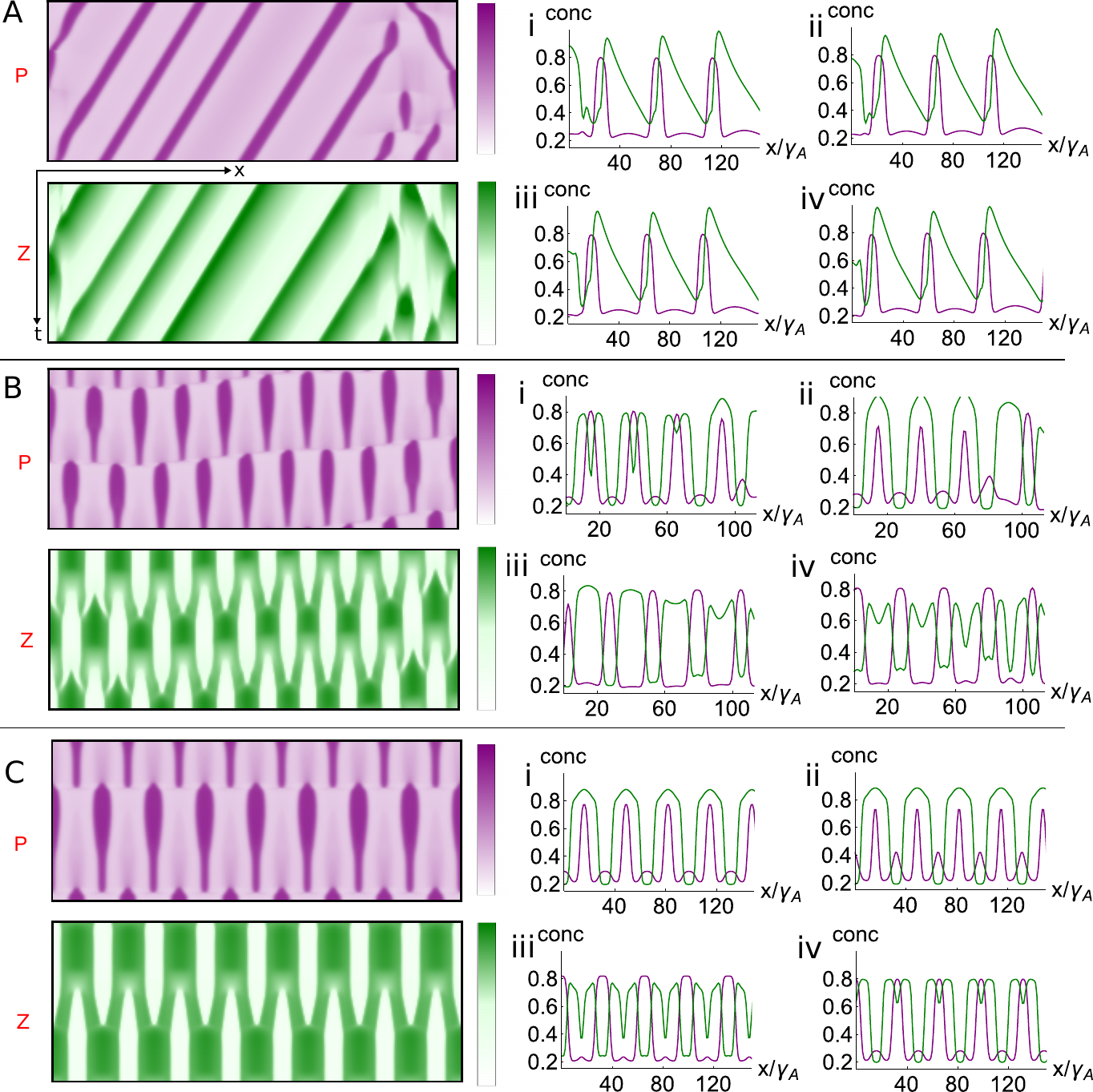}
    \caption{Dynamics that can be obtained when two phase separating species (model (M2)) interact chemically through the autocatalytic  reactions~\eqref{eq:C1}. These dynamics can be classified as  (A) travelling pulses, (B)``cascading" pulses or (C) standing pulses. In all cases, sustained dynamics appear to be critically dependent on the fact that colocalization of $P$ and $Z$ is disfavored ($\chi>0$), and by mismatched diffusion coefficients ($\delta\not = 1$).}
    \label{fig:fig1}
\end{figure}

\begin{figure*}[htbp]
    \includegraphics[width=180mm]{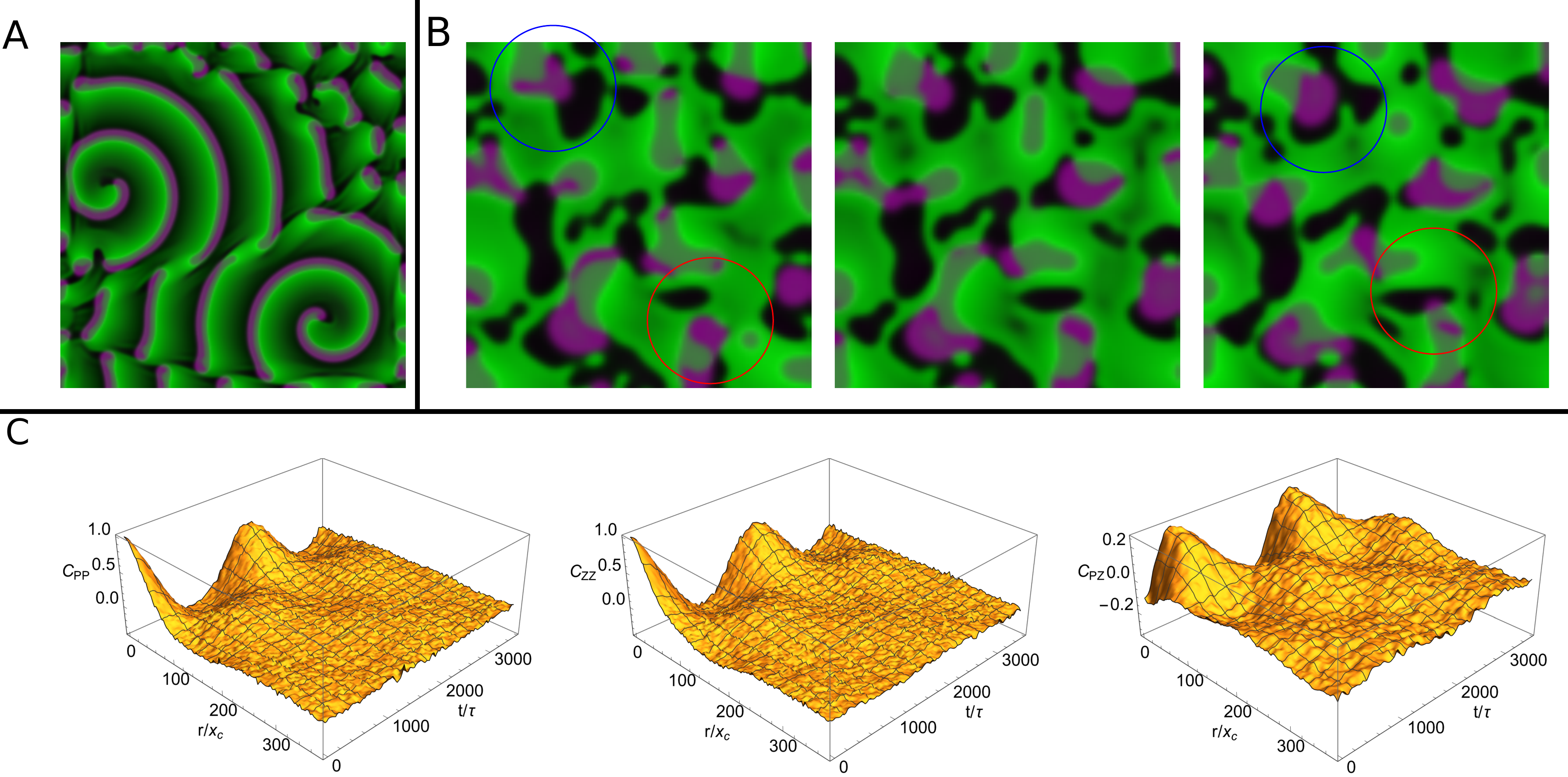}
    \caption{The different regimes when translated to two dimensions, showing $P$ (Purple) and $Z$ (Green) A) Travelling waves in one dimension tend to form spiral waves in two dimensions. B) Standing pulse phase tends to produce chaotic droplets in two dimensions. The droplets move into the gaps left by the phase separation of the $Z$ (Blue circle). Pulses are also born in regions of low concentration of $Z$ (red circle).}
    \label{fig:fig2}
\end{figure*}

\subsection{Emergence of oscillatory patterns}
Here we computationally examine the solutions to the full models (M1) and (M2), including reactions~\eqref{eq:C1}. We chose parameters so that oscillations are predicted near equilibrium, based on the fact that conditions~\eqref{eq:cond1}-\eqref{eq:cond4} are satisfied. This will allow us to get intuition about which specific parameters influence the emergence of sustained dynamics far from equilibrium. All the parameters used in our simulations are listed in Table I in the SI file.

{\bf Model (M2):} We begin by discussing the case in which both species can phase separate: model~\eqref{eq:M2} coupled with reactions~\eqref{eq:C1}. This model provides a much richer set of behaviors when compared to model~\eqref{eq:M1}, which is discussed at the end of this section. 

We first examine the solutions in a one dimensional space.  Fig.~\ref{fig:fig1} includes representative examples of the different forms of dynamics that we observed.

The spatio-temporal dynamics of the system depend most sensitively on the choice of  parameters $\chi$, $\delta$, $g$ in the model. Our exhaustive searches through these parameters and phase separation propensity  yield dynamics that can be categorized into one of these three classes: travelling pulses (Fig.~\ref{fig:fig1}A), cascading pulses (Fig.~\ref{fig:fig1}B), and standing pulses (Fig.~\ref{fig:fig1}C).  As the name suggests, travelling pulses correspond to spikes of $P$ density that spatially translate at a constant rate. Standing pulses are similar to standing waves, as pulses form and disappear at defined positions in space. Cascading pulses can be considered an intermediate behavior, in which standing pulses appear, but not synchronously: the emergence of a pulse at a particular location is temporally followed by the emergence of another pulse at a neighboring location (we shall discuss this in more depth later). See SI Figure 6 for phase diagram of transitions in these states.

 We found that changes in the diffusion ratio $\delta$ majorly influence the system's dynamic outcome. In general, travelling pulses appear for small values of $\delta$, standing pulses begin to appear as $\delta$ is increased, and at large $\delta$, the system doesn't exhibit any sustained dynamics. Additional plots summarizing the influence of parameter $g$,  $\delta$, and of the chemical reaction terms on the dynamic behaviors are reported SI Section V.
 
The emergence of such different outcomes in our phase separating system can be explained by some simple physical interpretations.  For $\delta<1$, species $P$  phase separates faster than $Z$. This leads to the formation of growing dense regions of $P$. In regions at high $P$ density, the autocatalytic reaction  promotes the conversion of $P$ into $Z$. Once the concentration of $Z$ passes a critical threshold, it is no longer preferable to maintain a high density of $P$ given the repulsive interaction energy between $Z$ and $P$, so the local peak of $P$ should disappear. If the diffusion coefficient of $Z$ is very small, this results in translation of the local peak of $P$ density, which manifests as the travelling pulse solutions that leave behind them a ``trail" of $Z$. If the diffusion coefficient of $Z$ is sufficiently large, the pulse P cannot ``escape" from Z, and instead this leads to a rapidly growing pulse that bursts up where the concentration of $Z$ is low and it is favorable for $P$ to phase separate again. 

The travelling waves observed in this system are comparable to those occurring in a reaction-diffusion context where $P$ and $Z$ do not undergo phase separation. The presence of phase separation increases complexity by promoting local increases in $Z$ and $P$ density, which achieves specific levels in the dense and in the dispersed phase based on the free energy~\eqref{eq:LGfunc}. This aspect  is especially important for the formation of the standing/cascading pulse dynamics, as we can observe that the maxima and minima of the pulses observed in Fig.~\ref{fig:fig1} correspond to the choices of the dense and dilute phase densities. All of these observed dynamical behaviors are persistent.
%\EF{A diagram of the transitions between these different behaviors can be found in SI Section XX.}

Even more diverse spatio-temporal behaviors occur when model~\eqref{eq:M2} is simulated in two dimensions. The same logic of formation of dynamics described above holds in two dimensions with a key difference, that is there are more than two directions in which spikes in $P$ can ``escape'' from regions dense in $Z$. This leads to the rich dynamics represented in Fig.~\ref{fig:fig2}. Travelling waves in one dimension become spiral waves in two dimensions (Fig.~\ref{fig:fig2}A), a standard dynamical pattern in reaction-diffusion systems (Supplementary Video V1). A less predictable outcome is that a system exhibiting standing pulses in one dimension turns into a system with disorganized dynamical phases in two dimensions, as can be observed in Fig.~\ref{fig:fig2}B Supplementary Video V2). We see here the same mechanism outlined previously in the formation of pulses, but when the process is embedded in higher dimensions there are more avenues for the ``escape" of $P$-dense droplets, corresponding to $P$-pulses in one dimension. If there are nearby regions at low density of $Z$ (caused by phase separation of $Z$), droplets of $P$ can translate into these spaces as exemplified in the blue circled part of Fig.~\ref{fig:fig2}B. If the concentration of $Z$ is consistently high nearby,  $P$ droplets tend to dissolve since they cannot translate to a neighboring $Z$-poor region, but new $P$ droplets can emerge in a distant $Z$-poor  area (red circled region). Therefore, the dynamics of the phase separating material $Z$ are crucial for the appearance of regions empty of $Z$ where  $P$ droplets can move into. Similarly to the one dimensional cases, the relative ratio of the diffusion constants $\delta$ is the most important factor in the appearance of these disordered droplets, as when the diffusion constant of $Z$ is too low they do not undergo phase separation dynamics quickly enough to leave a space for the spike of $P$ to move into. If they are too fast, however, no dynamics occur at all, as they relax faster than the spikes of $P$, which ``fill in the gaps" up to a chemical equilibrium. We note that these dynamics are persistent in time and occur even if the system is instantiated in a spiral conformation (see SI Fig. 7). These dynamics are persistent over long simulation times.

We quantitatively evaluated the structures observed in our simulations by computing the pair correlation function between concentrations at two different points in space and time~\cite{castelino_spatiotemporal_2020}:
\begin{equation}
C_{ij}(r,\tau)=\langle c_i(\mathbf x+r(\cos\theta,\sin\theta),t+\tau)c_j(\mathbf x,t)\rangle,
\end{equation}
where $c_i$ and $c_j$ represent the concentrations of chemical species $i$ and $j$; $x$ represents a point in space; ($r$, $\theta$) represent polar coordinates of points neighboring $x$; and finally $t$, $t+\tau$ are distinct times. The average is taken over $t,\mathbf x,\theta$.

The correlation functions for the concentration of $P$ and $Z$ under model~\eqref{eq:M2} is shown in Fig.~\ref{fig:fig3}A. We can see that spatial correlations in the concentrations of $P$ decay almost immediately. The structure of the temporal correlations appear to show several peaks. While droplets have local structure, as can be observed in the decay of the first peak, beyond their immediate vicinity correlations are lost very rapidly, indicating lack of information as to the structure over longer length scales. The correlation functions $C_{\text{PP}}$ and $C_{\text{ZZ}}$ have identical structures. In contrast, $C_{\text{PZ}}$ takes negative values as $r\to 0,t\to0$, reflective of the fact that mixing of $P$ and $Z$ is disfavored in our model.

{\bf Model (M1):} A contrast exists to  model~\eqref{eq:M1}, where only one of the two species phase separates. Extensive numerical simulations of this model in one dimension do not yield the same standing pulse solutions of the case when both species phase separate (SI Fig. 4 and SI Fig 5.). Therefore, when looked at in two dimensions, we do not find the disordered droplets as in model~\eqref{eq:M2}. While this does not constitute mathematical proof of the impossibility of such states in model~\eqref{eq:M1}, it is evidence that spatio-temporal disorder in phase separating systems is easier to achieve when there are multiple phase separating species. The emergence of the disordered droplets in model~\eqref{eq:M2} appears to depend strongly on low concentrations of $Z$ existing in certain areas because $Z$ phase separates.

\begin{figure}[htbp]
\includegraphics[width=\columnwidth]{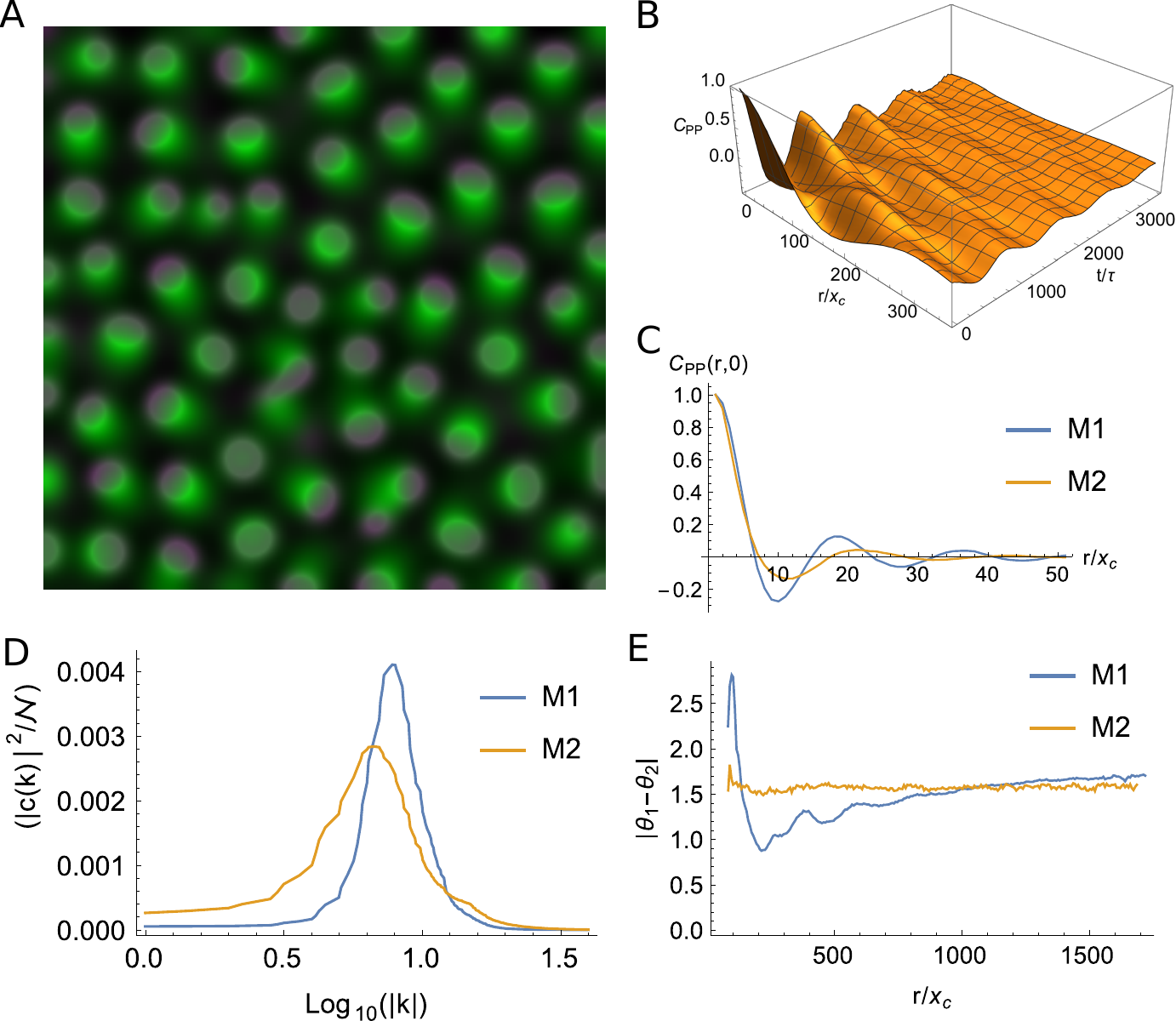}
    \caption{When one of the species is no longer phase separating (model M2), the spatial patterning becomes more organized. A) The purple droplets $P$ are driven forward by $Z$. B) The spatiotemporal correlations corresponding to A) show more peaks, indicating a greater degree of spatiotemporal order. C) Comparisons in space between model (M1) from figure \ref{fig:fig3} and model (M2). M2 has more peaks, and the peaks are higher. D) The (normalized) average power spectrum of concentration. Model M2 tends to have broader power spectra. E) The average direction between droplets as a function of distance between droplets. Model M1 tends to have more coherent motion. }
    \label{fig:fig4}
\end{figure}

Searching through parameters in model~\eqref{eq:M1} does yield dynamical droplets, though (after relaxation) they have regular patterning that change coherently in time. Example simulations are in Fig.~\ref{fig:fig4}A and  Supplementary Video V3. 
As before, the density of $Z$ ``drives" forward $P$-dense droplets because mixing of $P$ and $Z$ is disfavored. However because $Z$ does not undergo any phase separation in the absence of $P$, the regions into which $P$ droplets travel do not have significant heterogeneity in the density of $Z$. For this reason, the system does not exhibit the irregular droplet patterning observed in model~\eqref{eq:M2}, and the occurrence of  droplet birth, death, fusion and splitting. Consistent with these qualitative observations, the density correlation function for model~\eqref{eq:M1} displays greater structure (more peaks) when compared to model~\eqref{eq:M2}, as illustrated in Fig.~\ref{fig:fig4}B and C. In particular, focusing on the slice $t=0$ in Fig.~\ref{fig:fig4}B, peaks of the correlation function $C_{\text{PP}}$ persist at larger distances, reflecting the appearance of longer range order.  Further information can be obtained by looking at the (Fourier) power spectrum in space (averaged over 1000 snapshots), as seen in Fig.~\ref{fig:fig4}D, where it can be seen that model ~\eqref{eq:M2} has a broader distribution over length scales  than (M1), indicating that both droplets and the distances between droplets have a broader distribution in ~\eqref{eq:M2}.  Finally, we can study how the droplets move in space by looking at the direction of motion of any particular droplet, which is given by:
\begin{equation}
\theta = \text{arctan}((y(t+\delta t)-y(t))/(x(t+\delta t)-x(t)))
\end{equation}

where $x,y$ are the centres of mass of a droplet. We study how the average between the directions droplets move in vary as a function of distances between droplets. This can be seen in Fig.~\ref{fig:fig4}E. Very close droplets in model ~\eqref{eq:M1} have a tendency to move towards one another (a difference of $\pi$ corresponds to two droplets moving in exactly opposite directions). However, above a certain length, the droplets have a tendency to move more in phase. As the distance increases the coherence of motion decreases. At long distances there is even a slight tendency to move in anti-correlated manner due to the presence of boundaries between regions of correlated motion as can be seen in Supplementary Video V3. In contrast, model ~\eqref{eq:M2} does not display any strong trends of coherent motion between droplets, indicating that droplets are mainly moving randomly.

\subsection{Interpreting complex droplet dynamics through discrete rules}
The previous section showed that model (M2) can produce droplets of different sizes, distributed over many length scales, with no phase coherence in their dynamics. Is there a simple way to understand these complex behaviours?

Because many processes coexist in our models (phase separation, chemical reactions, diffusion, surface effects etc.), the analysis of the resultant dynamics is quite complicated, especially as a function of the many parameters needed to specify even a minimal model. However, these  dynamics may be understood through an analogy to alternative, more abstract systems. For example, the appearance of disordered droplet dynamics is reminiscent of cellular automata~\cite{adamatzky2010game}. As illustrated in Supplementary Video V1, the ``complex'' droplets exhibit an array of behaviors in which we can recognize droplet birth, death, splitting, fusion and motion. Moving to a level of abstraction beyond the PDEs defined in~\eqref{eq:fund1}-\eqref{eq:fund2}, we can treat the system as an approximate lattice where sites are either occupied or unoccupied according to associated value of concentration. We define the \emph{Occupancy} of a species having concentration $c$ a particular point in space $(x,y)$:
\begin{equation}
    O(x,y)=\begin{cases}
    1 \text{ if } c(x,y)> \frac{1}{2}(\rho_1+\rho_0)  \\ 
    0 \text{ if } c(x,y)< \frac{1}{2}(\rho_1+\rho_0), 
    \end{cases}
\end{equation}
Where $\rho_1$ and $\rho_0$ are the dense and dispersed concentrations in the general free energy model~\eqref{eq:LGfunc}.  Phase separating systems are particularly amenable to this kind of abstraction, precisely because the underlying free energy models thermodynamically favor equilibrium concentration close to specific values $\rho_0$ and $\rho_1$ for the phase separating species (i.e., to first approximation occupancy 1 states will have a concentration of $\rho_1$ and occupancy 0 states will have a concentration of $\rho_0$). This is reminiscent of binary ``on'' or ``off'' states typically associated with automata.

\begin{figure*}[htpb]
    \includegraphics[width=180mm]{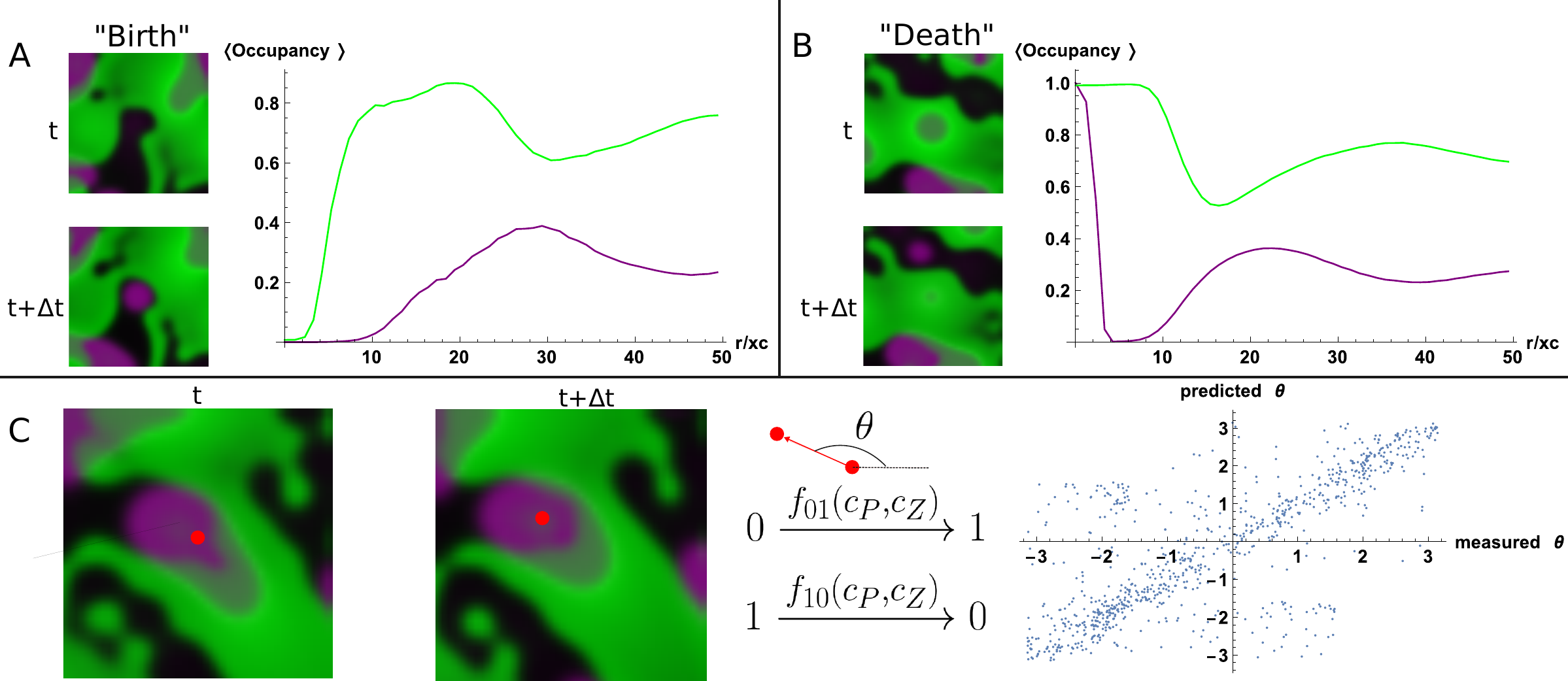}
    \caption{A) The ``birth" of new droplets arises from regions which are low in both concentrations of $Z$ without any other $P$ droplets around. This leads to agglomeration of $P$ that is escaping the $Z$ concentration. Mean occupancy (defined in the text) of $P$ and $Z$ captures this phenomena, showing the local structure around regions where droplets are born. B) ``Death of droplets" If a $P$ droplet finds itself immersed in concentration of $Z$, it can disappear as it has nowhere to translate into. The mean occupancy again shows that droplets tend to die if surrounded by $Z$.  C) $P$ droplets move into the gaps left by the phase separation of the concentration of $Z$. This leads to apparent motion. Analyzing the centre of mass of the droplets (red drop) over time and measuring the angle of motion (in the lab frame) allows us to predict motion using a simple cellular automata scheme of droplet dynamics, which is broadly in agreement with the actual measured motion of the droplets.}
    \label{fig:fig3}
\end{figure*}

When looking at particular kinds of events, such as the birth of droplets or their death, we can see how the local structure of occupancy of both $P$ and $Z$ is  related to such events. We can study the occupancy of all the snapshots of our system around the points droplets appear or disappear. Fig.~\ref{fig:fig3}B illustrates clearly that $P$ droplets tend to appear in regions where the concentration of $Z$ is low, and is relatively far from other droplets of $P$. ``Birth'' of  $P$ droplets can occur where $P$ is not being suppressed by $Z$. Conversely, ``death" of $P$ droplets occurs when they are are immersed in high $Z$ areas. 

The fact that occupancy is predictive of subsequent behavior enables us to consider whether we could define a ``rule set" as exists in cellular automata to predict how droplets behave after a single timestep. These rules can be discerned empirically, sharing the features of classic cellular automata rulesets where we have transitions between states ``0" and ``1" that are governed by the local environment. We propose these approximate rules to describe the temporal evolution of $P$ at a site characterized by $O_P(x,y)$ and $O_Z(x,y)$:

\begin{align}
& 1\to 0 \text{ if } O_P=1,O_Z=1,\rho_P<\rho_c \\
& 1\to 1 \text{ if } O_P=1,O_Z=1,\rho_P>\rho_c \\
& 0\to 1 \text{ if } O_P=0,O_Z=0,\rho_P>\rho_c \\
& 0\to 0 \text{ if } O_P=0,O_Z=0,\rho_P<\rho_c \\
& 1\to 1 \text{ if } O_P=1,O_Z=0, \\
& 0\to 0 \text{ if } O_P=0,O_Z=0
\end{align}

These rules take into account the concentration of neighboring sites, $\rho_P$ (up to some length scale), and compare them to an empirically determined value of $\rho_c$ that represents a critical concentration leading to loss or gain of $P$ at the site. These simple rules already lead to birth and death of new droplets depending on the local structure, but they can also predict how the droplets move in time, the $\theta$ quantity we measured in the previous section, which did not display any coherent droplet motion. We can compare solutions of the PDE to the cellular automata scheme and show that even this very pared down form of dynamics can predict what happens to the time evolution of $P$ droplets. The difference in predicted versus observed $\theta$, which is the direction of motion, can be observed in Fig.~\ref{fig:fig3}C. The short-time prediction as to the subsequent motion of the droplet is well captured by our elementary cellular automaton. From this, we can see that our droplets effectively compete with one another for space, while also producing $Z$ that is harmful to other droplets.

As these rules continue to be iterated forward for the occupancy of $P$, a steady state arises, unlike in the full PDE system, as we neglected the time evolution of the more slowly diffusing $Z$. This illustrates that spatial relaxation across different timescales is necessary for continual dynamics to arise. Our recommendation for disordered dynamics can thus be summarized as ``fast droplets with slow environmental reconfiguration".

\section{Discussion and Conclusion}

In this study, we have examined reaction-phase separation systems including two species, discussing the emergence of spatio-temporal dynamics through a linearized approach. We then analyzed computationally a particular system that includes autocatalytic reactions, and mapped its possible dynamical behaviors in both one and two dimensions. We found that when both species in our model phase separate, there exists parameter regimes where they  ``standing pulse" phases appear in one dimension. In two dimensions, this scenario corresponds to droplets continually moving through space, with a very short correlation length in the concentration correlation function, implying a lack of spatial order but the presence of temporal order. This dynamical regime appears to require both species to phase separate, because similar dynamics are not observed when only one species phase separates and the other simply diffuses in space. Our observations indicate that the spatial heterogeneity induced by phase separation is a key factor responsible for the emergence of dynamic, spatially disordered states.

We have contented ourselves with mainly simulations and linear stability of the homogeneous state in this work. Unfortunately, this means that we cannot be as categorical with the possibilities that exist only in systems with phase separation. Not only due to the proliferation of parameters in this system but possibly with different forms of chemical dynamics or more species. Thus far, we demonstrated the emergence of complex dynamics in phase separating systems where both species phase separate, but this does not constitute a proof that such behavior isn't possible in the absence of phase separation. We swept parameter space when one of the species wasn't phase separating and could not find similar dynamical behaviors, however we cannot exclude that such dynamics could exist. Further work will have to incorporate analytical approaches to demonstrate the full richness of dynamics that can exist with phase separation. It is possible that extending the system to more components could reproduce these behaviors, as chemical reaction networks containing bistability can resemble the tendency of phase separating systems to form at two specific densities.

Some of the behaviors we observed are reminiscent of spatio-temporal chaos observed in reaction-diffusion systems~\cite{zhang_chaotic_1995,rossler_chemical_1976,coullet_defect-mediated_1989,epstein2016reaction}. However, the presence of phase separation and surface tension leads to the tendency for these waves fronts to be concentrated into pulses, leading to the travelling droplet phases we observe. The dynamics we observe are similar to the dynamics seen in other models of excitable media~\cite{greenberg1978spatial}. In contrast to standard excitable media, however, in phase separating systems the homogeneous phase is not usually linearly stable to perturbations. 

Reaction-diffusion schemes have been a canonical model in development, and recent approaches have considered cellular automata like schemes towards the explanation of patterning in organisms~\cite{zakany_simple_2023}. Separately, phase separation has been the object of recent study both as a paradigm towards understanding early life and for its role in intracellular regulation~\cite{boeynaems2018protein,keating2012aqueous}. In this work, we have reconsidered phase separating systems as a continuum model of cellular automata. Cellular automata have long been of interest to the mathematical community as a simple paradigm for the emergence of complexity from simple rulesets~\cite{adamatzky2010game}. Similarly, droplets that can be programmed to interact with one another chemically can generate effective ``rulesets'' that lead to complex behaviors, as seen in this study. The full possibilities of emerging states in such a paradigm has yet to be fully characterized.  

\bibliography{references}

%merlin.mbs apsrev4-1.bst 2010-07-25 4.21a (PWD, AO, DPC) hacked
%Control: key (0)
%Control: author (8) initials jnrlst
%Control: editor formatted (1) identically to author
%Control: production of article title (-1) disabled
%Control: page (0) single
%Control: year (1) truncated
%Control: production of eprint (0) enabled
\begin{thebibliography}{40}%
\makeatletter
\providecommand \@ifxundefined [1]{%
 \@ifx{#1\undefined}
}%
\providecommand \@ifnum [1]{%
 \ifnum #1\expandafter \@firstoftwo
 \else \expandafter \@secondoftwo
 \fi
}%
\providecommand \@ifx [1]{%
 \ifx #1\expandafter \@firstoftwo
 \else \expandafter \@secondoftwo
 \fi
}%
\providecommand \natexlab [1]{#1}%
\providecommand \enquote  [1]{``#1''}%
\providecommand \bibnamefont  [1]{#1}%
\providecommand \bibfnamefont [1]{#1}%
\providecommand \citenamefont [1]{#1}%
\providecommand \href@noop [0]{\@secondoftwo}%
\providecommand \href [0]{\begingroup \@sanitize@url \@href}%
\providecommand \@href[1]{\@@startlink{#1}\@@href}%
\providecommand \@@href[1]{\endgroup#1\@@endlink}%
\providecommand \@sanitize@url [0]{\catcode `\\12\catcode `\$12\catcode
  `\&12\catcode `\#12\catcode `\^12\catcode `\_12\catcode `\%12\relax}%
\providecommand \@@startlink[1]{}%
\providecommand \@@endlink[0]{}%
\providecommand \url  [0]{\begingroup\@sanitize@url \@url }%
\providecommand \@url [1]{\endgroup\@href {#1}{\urlprefix }}%
\providecommand \urlprefix  [0]{URL }%
\providecommand \Eprint [0]{\href }%
\providecommand \doibase [0]{http://dx.doi.org/}%
\providecommand \selectlanguage [0]{\@gobble}%
\providecommand \bibinfo  [0]{\@secondoftwo}%
\providecommand \bibfield  [0]{\@secondoftwo}%
\providecommand \translation [1]{[#1]}%
\providecommand \BibitemOpen [0]{}%
\providecommand \bibitemStop [0]{}%
\providecommand \bibitemNoStop [0]{.\EOS\space}%
\providecommand \EOS [0]{\spacefactor3000\relax}%
\providecommand \BibitemShut  [1]{\csname bibitem#1\endcsname}%
\let\auto@bib@innerbib\@empty
%</preamble>
\bibitem [{\citenamefont {Rai}\ \emph {et~al.}(2018)\citenamefont {Rai},
  \citenamefont {Chen}, \citenamefont {Selbach},\ and\ \citenamefont
  {Pelkmans}}]{rai2018kinase}%
  \BibitemOpen
  \bibfield  {author} {\bibinfo {author} {\bibfnamefont {A.~K.}\ \bibnamefont
  {Rai}}, \bibinfo {author} {\bibfnamefont {J.-X.}\ \bibnamefont {Chen}},
  \bibinfo {author} {\bibfnamefont {M.}~\bibnamefont {Selbach}}, \ and\
  \bibinfo {author} {\bibfnamefont {L.}~\bibnamefont {Pelkmans}},\ }\href@noop
  {} {\bibfield  {journal} {\bibinfo  {journal} {Nature}\ }\textbf {\bibinfo
  {volume} {559}},\ \bibinfo {pages} {211} (\bibinfo {year}
  {2018})}\BibitemShut {NoStop}%
\bibitem [{\citenamefont {Garabedian}\ \emph {et~al.}(2021)\citenamefont
  {Garabedian}, \citenamefont {Wang}, \citenamefont {Dabdoub}, \citenamefont
  {Tong}, \citenamefont {Caldwell}, \citenamefont {Benman}, \citenamefont
  {Schuster}, \citenamefont {Deiters},\ and\ \citenamefont
  {Good}}]{garabedian2021designer}%
  \BibitemOpen
  \bibfield  {author} {\bibinfo {author} {\bibfnamefont {M.~V.}\ \bibnamefont
  {Garabedian}}, \bibinfo {author} {\bibfnamefont {W.}~\bibnamefont {Wang}},
  \bibinfo {author} {\bibfnamefont {J.~B.}\ \bibnamefont {Dabdoub}}, \bibinfo
  {author} {\bibfnamefont {M.}~\bibnamefont {Tong}}, \bibinfo {author}
  {\bibfnamefont {R.~M.}\ \bibnamefont {Caldwell}}, \bibinfo {author}
  {\bibfnamefont {W.}~\bibnamefont {Benman}}, \bibinfo {author} {\bibfnamefont
  {B.~S.}\ \bibnamefont {Schuster}}, \bibinfo {author} {\bibfnamefont
  {A.}~\bibnamefont {Deiters}}, \ and\ \bibinfo {author} {\bibfnamefont
  {M.~C.}\ \bibnamefont {Good}},\ }\href@noop {} {\bibfield  {journal}
  {\bibinfo  {journal} {Nature chemical biology}\ }\textbf {\bibinfo {volume}
  {17}},\ \bibinfo {pages} {998} (\bibinfo {year} {2021})}\BibitemShut
  {NoStop}%
\bibitem [{\citenamefont {Agarwal}\ \emph {et~al.}(2024)\citenamefont
  {Agarwal}, \citenamefont {Osmanovic}, \citenamefont {Dizani}, \citenamefont
  {Klocke},\ and\ \citenamefont {Franco}}]{agarwal2024dynamic}%
  \BibitemOpen
  \bibfield  {author} {\bibinfo {author} {\bibfnamefont {S.}~\bibnamefont
  {Agarwal}}, \bibinfo {author} {\bibfnamefont {D.}~\bibnamefont {Osmanovic}},
  \bibinfo {author} {\bibfnamefont {M.}~\bibnamefont {Dizani}}, \bibinfo
  {author} {\bibfnamefont {M.~A.}\ \bibnamefont {Klocke}}, \ and\ \bibinfo
  {author} {\bibfnamefont {E.}~\bibnamefont {Franco}},\ }\href@noop {}
  {\bibfield  {journal} {\bibinfo  {journal} {Nature Communications}\ }\textbf
  {\bibinfo {volume} {15}},\ \bibinfo {pages} {1915} (\bibinfo {year}
  {2024})}\BibitemShut {NoStop}%
\bibitem [{\citenamefont {Deng}\ and\ \citenamefont
  {Walther}(2020)}]{deng2020programmable}%
  \BibitemOpen
  \bibfield  {author} {\bibinfo {author} {\bibfnamefont {J.}~\bibnamefont
  {Deng}}\ and\ \bibinfo {author} {\bibfnamefont {A.}~\bibnamefont {Walther}},\
  }\href@noop {} {\bibfield  {journal} {\bibinfo  {journal} {Chem}\ }\textbf
  {\bibinfo {volume} {6}},\ \bibinfo {pages} {3329} (\bibinfo {year}
  {2020})}\BibitemShut {NoStop}%
\bibitem [{\citenamefont {Ringkvist}\ and\ \citenamefont
  {Zhou}(2009)}]{ringkvist_dynamical_2009}%
  \BibitemOpen
  \bibfield  {author} {\bibinfo {author} {\bibfnamefont {M.}~\bibnamefont
  {Ringkvist}}\ and\ \bibinfo {author} {\bibfnamefont {Y.}~\bibnamefont
  {Zhou}},\ }\href {\doibase 10.1016/j.na.2009.01.149} {\bibfield  {journal}
  {\bibinfo  {journal} {Nonlinear Analysis: Theory, Methods \& Applications}\
  }\textbf {\bibinfo {volume} {71}},\ \bibinfo {pages} {2667} (\bibinfo {year}
  {2009})}\BibitemShut {NoStop}%
\bibitem [{\citenamefont {Zheng}\ and\ \citenamefont
  {Shen}(2015)}]{zheng_pattern_2015}%
  \BibitemOpen
  \bibfield  {author} {\bibinfo {author} {\bibfnamefont {Q.}~\bibnamefont
  {Zheng}}\ and\ \bibinfo {author} {\bibfnamefont {J.}~\bibnamefont {Shen}},\
  }\href {\doibase 10.1016/j.camwa.2015.06.031} {\bibfield  {journal} {\bibinfo
   {journal} {Computers \& Mathematics with Applications}\ }\textbf {\bibinfo
  {volume} {70}},\ \bibinfo {pages} {1082} (\bibinfo {year}
  {2015})}\BibitemShut {NoStop}%
\bibitem [{\citenamefont {McGough}\ and\ \citenamefont
  {Riley}(2004)}]{mcgough_pattern_2004}%
  \BibitemOpen
  \bibfield  {author} {\bibinfo {author} {\bibfnamefont {J.~S.}\ \bibnamefont
  {McGough}}\ and\ \bibinfo {author} {\bibfnamefont {K.}~\bibnamefont
  {Riley}},\ }\href {\doibase 10.1016/S1468-1218(03)00020-8} {\bibfield
  {journal} {\bibinfo  {journal} {Nonlinear Analysis: Real World Applications}\
  }\textbf {\bibinfo {volume} {5}},\ \bibinfo {pages} {105} (\bibinfo {year}
  {2004})}\BibitemShut {NoStop}%
\bibitem [{\citenamefont {Kirschbaum}\ and\ \citenamefont
  {Zwicker}(2021)}]{kirschbaum2021controlling}%
  \BibitemOpen
  \bibfield  {author} {\bibinfo {author} {\bibfnamefont {J.}~\bibnamefont
  {Kirschbaum}}\ and\ \bibinfo {author} {\bibfnamefont {D.}~\bibnamefont
  {Zwicker}},\ }\href@noop {} {\bibfield  {journal} {\bibinfo  {journal}
  {Journal of The Royal Society Interface}\ }\textbf {\bibinfo {volume} {18}},\
  \bibinfo {pages} {20210255} (\bibinfo {year} {2021})}\BibitemShut {NoStop}%
\bibitem [{\citenamefont {Osmanovi{\'{c}}}\ and\ \citenamefont
  {Franco}(2023)}]{Osmanovic2023}%
  \BibitemOpen
  \bibfield  {author} {\bibinfo {author} {\bibfnamefont {D.}~\bibnamefont
  {Osmanovi{\'{c}}}}\ and\ \bibinfo {author} {\bibfnamefont {E.}~\bibnamefont
  {Franco}},\ }\href {\doibase 10.1098/rsif.2023.0117} {\bibfield  {journal}
  {\bibinfo  {journal} {Journal of The Royal Society Interface}\ }\textbf
  {\bibinfo {volume} {20}},\ \bibinfo {pages} {20230117} (\bibinfo {year}
  {2023})},\ \Eprint
  {http://arxiv.org/abs/https://royalsocietypublishing.org/doi/pdf/10.1098/rsif.2023.0117}
  {https://royalsocietypublishing.org/doi/pdf/10.1098/rsif.2023.0117}
  \BibitemShut {NoStop}%
\bibitem [{\citenamefont {Luo}\ and\ \citenamefont
  {Zwicker}(2023)}]{luo2023influence}%
  \BibitemOpen
  \bibfield  {author} {\bibinfo {author} {\bibfnamefont {C.}~\bibnamefont
  {Luo}}\ and\ \bibinfo {author} {\bibfnamefont {D.}~\bibnamefont {Zwicker}},\
  }\href@noop {} {\bibfield  {journal} {\bibinfo  {journal} {Physical Review
  E}\ }\textbf {\bibinfo {volume} {108}},\ \bibinfo {pages} {034206} (\bibinfo
  {year} {2023})}\BibitemShut {NoStop}%
\bibitem [{\citenamefont {Bauermann}\ \emph {et~al.}(2023)\citenamefont
  {Bauermann}, \citenamefont {Bartolucci}, \citenamefont {Boekhoven},
  \citenamefont {Weber},\ and\ \citenamefont
  {Jülicher}}]{bauermann_formation_2023}%
  \BibitemOpen
  \bibfield  {author} {\bibinfo {author} {\bibfnamefont {J.}~\bibnamefont
  {Bauermann}}, \bibinfo {author} {\bibfnamefont {G.}~\bibnamefont
  {Bartolucci}}, \bibinfo {author} {\bibfnamefont {J.}~\bibnamefont
  {Boekhoven}}, \bibinfo {author} {\bibfnamefont {C.~A.}\ \bibnamefont
  {Weber}}, \ and\ \bibinfo {author} {\bibfnamefont {F.}~\bibnamefont
  {Jülicher}},\ }\href {\doibase 10.1103/PhysRevResearch.5.043246} {\bibfield
  {journal} {\bibinfo  {journal} {Physical Review Research}\ }\textbf {\bibinfo
  {volume} {5}},\ \bibinfo {pages} {043246} (\bibinfo {year} {2023})},\
  \bibinfo {note} {publisher: American Physical Society}\BibitemShut {NoStop}%
\bibitem [{\citenamefont {Cho}\ and\ \citenamefont
  {Jacobs}(2023)}]{cho_nonequilibrium_2023}%
  \BibitemOpen
  \bibfield  {author} {\bibinfo {author} {\bibfnamefont {Y.}~\bibnamefont
  {Cho}}\ and\ \bibinfo {author} {\bibfnamefont {W.~M.}\ \bibnamefont
  {Jacobs}},\ }\href {\doibase 10.1063/5.0166824} {\bibfield  {journal}
  {\bibinfo  {journal} {The Journal of Chemical Physics}\ }\textbf {\bibinfo
  {volume} {159}},\ \bibinfo {pages} {154101} (\bibinfo {year}
  {2023})}\BibitemShut {NoStop}%
\bibitem [{\citenamefont {Zwicker}\ \emph {et~al.}(2016)\citenamefont
  {Zwicker}, \citenamefont {Seyboldt}, \citenamefont {Weber}, \citenamefont
  {Hyman},\ and\ \citenamefont {Jülicher}}]{zwicker_2016}%
  \BibitemOpen
  \bibfield  {author} {\bibinfo {author} {\bibfnamefont {D.}~\bibnamefont
  {Zwicker}}, \bibinfo {author} {\bibfnamefont {R.}~\bibnamefont {Seyboldt}},
  \bibinfo {author} {\bibfnamefont {C.~A.}\ \bibnamefont {Weber}}, \bibinfo
  {author} {\bibfnamefont {A.~A.}\ \bibnamefont {Hyman}}, \ and\ \bibinfo
  {author} {\bibfnamefont {F.}~\bibnamefont {Jülicher}},\ }\href {\doibase
  10.1038/nphys3984} {\bibfield  {journal} {\bibinfo  {journal} {Nature
  physics}\ }\textbf {\bibinfo {volume} {13}},\ \bibinfo {pages} {408}
  (\bibinfo {year} {2016})}\BibitemShut {NoStop}%
\bibitem [{\citenamefont {Li}\ and\ \citenamefont {Cates}(2020)}]{Li2020}%
  \BibitemOpen
  \bibfield  {author} {\bibinfo {author} {\bibfnamefont {Y.~I.}\ \bibnamefont
  {Li}}\ and\ \bibinfo {author} {\bibfnamefont {M.~E.}\ \bibnamefont {Cates}},\
  }\href {\doibase 10.1088/1742-5468/ab7e2d} {\bibfield  {journal} {\bibinfo
  {journal} {Journal of Statistical Mechanics: Theory and Experiment}\ }\textbf
  {\bibinfo {volume} {2020}} (\bibinfo {year} {2020}),\
  10.1088/1742-5468/ab7e2d},\ \bibinfo {note} {arXiv: 2001.02563 Publisher: IOP
  Publishing and SISSA}\BibitemShut {NoStop}%
\bibitem [{\citenamefont {Pearson}(1993)}]{pearson_complex_1993}%
  \BibitemOpen
  \bibfield  {author} {\bibinfo {author} {\bibfnamefont {J.~E.}\ \bibnamefont
  {Pearson}},\ }\href {\doibase 10.1126/science.261.5118.189} {\bibfield
  {journal} {\bibinfo  {journal} {Science}\ }\textbf {\bibinfo {volume}
  {261}},\ \bibinfo {pages} {189} (\bibinfo {year} {1993})},\ \bibinfo {note}
  {publisher: American Association for the Advancement of Science}\BibitemShut
  {NoStop}%
\bibitem [{\citenamefont {Lee}\ and\ \citenamefont
  {Swinney}(1995)}]{lee_lamellar_1995}%
  \BibitemOpen
  \bibfield  {author} {\bibinfo {author} {\bibfnamefont {K.~J.}\ \bibnamefont
  {Lee}}\ and\ \bibinfo {author} {\bibfnamefont {H.~L.}\ \bibnamefont
  {Swinney}},\ }\href {\doibase 10.1103/PhysRevE.51.1899} {\bibfield  {journal}
  {\bibinfo  {journal} {Physical Review E}\ }\textbf {\bibinfo {volume} {51}},\
  \bibinfo {pages} {1899} (\bibinfo {year} {1995})}\BibitemShut {NoStop}%
\bibitem [{\citenamefont {Demarchi}\ \emph {et~al.}(2023)\citenamefont
  {Demarchi}, \citenamefont {Goychuk}, \citenamefont {Maryshev},\ and\
  \citenamefont {Frey}}]{demarchi_enzyme-enriched_2023}%
  \BibitemOpen
  \bibfield  {author} {\bibinfo {author} {\bibfnamefont {L.}~\bibnamefont
  {Demarchi}}, \bibinfo {author} {\bibfnamefont {A.}~\bibnamefont {Goychuk}},
  \bibinfo {author} {\bibfnamefont {I.}~\bibnamefont {Maryshev}}, \ and\
  \bibinfo {author} {\bibfnamefont {E.}~\bibnamefont {Frey}},\ }\href {\doibase
  10.1103/PhysRevLett.130.128401} {\bibfield  {journal} {\bibinfo  {journal}
  {Physical Review Letters}\ }\textbf {\bibinfo {volume} {130}},\ \bibinfo
  {pages} {128401} (\bibinfo {year} {2023})},\ \bibinfo {note} {publisher:
  American Physical Society}\BibitemShut {NoStop}%
\bibitem [{\citenamefont {Zwicker}(2023)}]{zwicker_droplets_2023}%
  \BibitemOpen
  \bibfield  {author} {\bibinfo {author} {\bibfnamefont {D.}~\bibnamefont
  {Zwicker}},\ }\href {\doibase 10.1103/PhysRevLett.130.128401} {\bibfield
  {journal} {\bibinfo  {journal} {Physics}\ }\textbf {\bibinfo {volume} {16}},\
  \bibinfo {pages} {45} (\bibinfo {year} {2023})},\ \bibinfo {note} {publisher:
  American Physical Society}\BibitemShut {NoStop}%
\bibitem [{\citenamefont {Rössler}(1976)}]{rossler_chemical_1976}%
  \BibitemOpen
  \bibfield  {author} {\bibinfo {author} {\bibfnamefont {O.~E.}\ \bibnamefont
  {Rössler}},\ }\href {\doibase 10.1515/zna-1976-1006} {\bibfield  {journal}
  {\bibinfo  {journal} {Zeitschrift für Naturforschung A}\ }\textbf {\bibinfo
  {volume} {31}},\ \bibinfo {pages} {1168} (\bibinfo {year} {1976})},\ \bibinfo
  {note} {publisher: De Gruyter}\BibitemShut {NoStop}%
\bibitem [{\citenamefont {Zimmermann}\ \emph {et~al.}(1997)\citenamefont
  {Zimmermann}, \citenamefont {Firle}, \citenamefont {Natiello}, \citenamefont
  {Hildebrand}, \citenamefont {Eiswirth}, \citenamefont {Bär}, \citenamefont
  {Bangia},\ and\ \citenamefont {Kevrekidis}}]{zimmermann_pulse_1997}%
  \BibitemOpen
  \bibfield  {author} {\bibinfo {author} {\bibfnamefont {M.~G.}\ \bibnamefont
  {Zimmermann}}, \bibinfo {author} {\bibfnamefont {S.~O.}\ \bibnamefont
  {Firle}}, \bibinfo {author} {\bibfnamefont {M.~A.}\ \bibnamefont {Natiello}},
  \bibinfo {author} {\bibfnamefont {M.}~\bibnamefont {Hildebrand}}, \bibinfo
  {author} {\bibfnamefont {M.}~\bibnamefont {Eiswirth}}, \bibinfo {author}
  {\bibfnamefont {M.}~\bibnamefont {Bär}}, \bibinfo {author} {\bibfnamefont
  {A.~K.}\ \bibnamefont {Bangia}}, \ and\ \bibinfo {author} {\bibfnamefont
  {I.~G.}\ \bibnamefont {Kevrekidis}},\ }\href {\doibase
  10.1016/S0167-2789(97)00112-7} {\bibfield  {journal} {\bibinfo  {journal}
  {Physica D: Nonlinear Phenomena}\ }\textbf {\bibinfo {volume} {110}},\
  \bibinfo {pages} {92} (\bibinfo {year} {1997})}\BibitemShut {NoStop}%
\bibitem [{\citenamefont {Strain}\ and\ \citenamefont
  {Greenside}(1998)}]{strain_size-dependent_1998}%
  \BibitemOpen
  \bibfield  {author} {\bibinfo {author} {\bibfnamefont {M.~C.}\ \bibnamefont
  {Strain}}\ and\ \bibinfo {author} {\bibfnamefont {H.~S.}\ \bibnamefont
  {Greenside}},\ }\href {\doibase 10.1103/PhysRevLett.80.2306} {\bibfield
  {journal} {\bibinfo  {journal} {Physical Review Letters}\ }\textbf {\bibinfo
  {volume} {80}},\ \bibinfo {pages} {2306} (\bibinfo {year} {1998})},\ \bibinfo
  {note} {publisher: American Physical Society}\BibitemShut {NoStop}%
\bibitem [{\citenamefont {Cai}\ \emph {et~al.}(2001)\citenamefont {Cai},
  \citenamefont {McLaughlin},\ and\ \citenamefont
  {Shatah}}]{cai_spatiotemporal_2001}%
  \BibitemOpen
  \bibfield  {author} {\bibinfo {author} {\bibfnamefont {D.}~\bibnamefont
  {Cai}}, \bibinfo {author} {\bibfnamefont {D.~W.}\ \bibnamefont {McLaughlin}},
  \ and\ \bibinfo {author} {\bibfnamefont {J.}~\bibnamefont {Shatah}},\ }\href
  {\doibase 10.1016/S0378-4754(00)00299-8} {\bibfield  {journal} {\bibinfo
  {journal} {Mathematics and Computers in Simulation}\ }\textbf {\bibinfo
  {volume} {55}},\ \bibinfo {pages} {329} (\bibinfo {year} {2001})}\BibitemShut
  {NoStop}%
\bibitem [{\citenamefont {Castelino}\ \emph {et~al.}(2020)\citenamefont
  {Castelino}, \citenamefont {Ratliff}, \citenamefont {Rucklidge},
  \citenamefont {Subramanian},\ and\ \citenamefont
  {Topaz}}]{castelino_spatiotemporal_2020}%
  \BibitemOpen
  \bibfield  {author} {\bibinfo {author} {\bibfnamefont {J.~K.}\ \bibnamefont
  {Castelino}}, \bibinfo {author} {\bibfnamefont {D.~J.}\ \bibnamefont
  {Ratliff}}, \bibinfo {author} {\bibfnamefont {A.~M.}\ \bibnamefont
  {Rucklidge}}, \bibinfo {author} {\bibfnamefont {P.}~\bibnamefont
  {Subramanian}}, \ and\ \bibinfo {author} {\bibfnamefont {C.~M.}\ \bibnamefont
  {Topaz}},\ }\href {\doibase 10.1016/j.physd.2020.132475} {\bibfield
  {journal} {\bibinfo  {journal} {Physica D: Nonlinear Phenomena}\ }\textbf
  {\bibinfo {volume} {409}},\ \bibinfo {pages} {132475} (\bibinfo {year}
  {2020})},\ \bibinfo {note} {arXiv:2001.11730 [nlin]}\BibitemShut {NoStop}%
\bibitem [{\citenamefont {Sayama}\ and\ \citenamefont
  {Nehaniv}(2024)}]{sayama_self-reproduction_2024}%
  \BibitemOpen
  \bibfield  {author} {\bibinfo {author} {\bibfnamefont {H.}~\bibnamefont
  {Sayama}}\ and\ \bibinfo {author} {\bibfnamefont {C.~L.}\ \bibnamefont
  {Nehaniv}},\ }\href {\doibase 10.48550/arXiv.2402.03961} {\enquote {\bibinfo
  {title} {Self-{Reproduction} and {Evolution} in {Cellular} {Automata}: 25
  {Years} after {Evoloops}},}\ } (\bibinfo {year} {2024}),\ \bibinfo {note}
  {arXiv:2402.03961 [nlin, q-bio]}\BibitemShut {NoStop}%
\bibitem [{\citenamefont {Hohenberg}\ and\ \citenamefont
  {Halperin}(1977)}]{Hohenberg1977}%
  \BibitemOpen
  \bibfield  {author} {\bibinfo {author} {\bibfnamefont {P.~C.}\ \bibnamefont
  {Hohenberg}}\ and\ \bibinfo {author} {\bibfnamefont {B.~I.}\ \bibnamefont
  {Halperin}},\ }\href {\doibase 10.1103/RevModPhys.49.435} {\bibfield
  {journal} {\bibinfo  {journal} {Reviews of Modern Physics}\ }\textbf
  {\bibinfo {volume} {49}},\ \bibinfo {pages} {435} (\bibinfo {year}
  {1977})}\BibitemShut {NoStop}%
\bibitem [{\citenamefont {Cross}\ and\ \citenamefont
  {Hohenberg}(1993)}]{Cross1993}%
  \BibitemOpen
  \bibfield  {author} {\bibinfo {author} {\bibfnamefont {M.~C.}\ \bibnamefont
  {Cross}}\ and\ \bibinfo {author} {\bibfnamefont {P.~C.}\ \bibnamefont
  {Hohenberg}},\ }\href {\doibase 10.1103/RevModPhys.65.851} {\bibfield
  {journal} {\bibinfo  {journal} {Reviews of Modern Physics}\ }\textbf
  {\bibinfo {volume} {65}},\ \bibinfo {pages} {851} (\bibinfo {year}
  {1993})}\BibitemShut {NoStop}%
\bibitem [{\citenamefont {Marsden}\ and\ \citenamefont
  {McCracken}(2012)}]{marsden2012hopf}%
  \BibitemOpen
  \bibfield  {author} {\bibinfo {author} {\bibfnamefont {J.~E.}\ \bibnamefont
  {Marsden}}\ and\ \bibinfo {author} {\bibfnamefont {M.}~\bibnamefont
  {McCracken}},\ }\href@noop {} {\emph {\bibinfo {title} {The {Hopf}
  bifurcation and its applications}}},\ Vol.~\bibinfo {volume} {19}\ (\bibinfo
  {publisher} {Springer Science \& Business Media},\ \bibinfo {year}
  {2012})\BibitemShut {NoStop}%
\bibitem [{\citenamefont {Gierer}\ and\ \citenamefont
  {Meinhardt}(1972)}]{gierer_theory_1972}%
  \BibitemOpen
  \bibfield  {author} {\bibinfo {author} {\bibfnamefont {A.}~\bibnamefont
  {Gierer}}\ and\ \bibinfo {author} {\bibfnamefont {H.}~\bibnamefont
  {Meinhardt}},\ }\href {\doibase 10.1007/BF00289234} {\bibfield  {journal}
  {\bibinfo  {journal} {Kybernetik}\ }\textbf {\bibinfo {volume} {12}},\
  \bibinfo {pages} {30} (\bibinfo {year} {1972})}\BibitemShut {NoStop}%
\bibitem [{\citenamefont {Landge}\ \emph {et~al.}(2020)\citenamefont {Landge},
  \citenamefont {Jordan}, \citenamefont {Diego},\ and\ \citenamefont
  {Müller}}]{landge_pattern_2020}%
  \BibitemOpen
  \bibfield  {author} {\bibinfo {author} {\bibfnamefont {A.~N.}\ \bibnamefont
  {Landge}}, \bibinfo {author} {\bibfnamefont {B.~M.}\ \bibnamefont {Jordan}},
  \bibinfo {author} {\bibfnamefont {X.}~\bibnamefont {Diego}}, \ and\ \bibinfo
  {author} {\bibfnamefont {P.}~\bibnamefont {Müller}},\ }\href {\doibase
  10.1016/j.ydbio.2019.10.031} {\bibfield  {journal} {\bibinfo  {journal}
  {Developmental Biology}\ }\bibinfo {series} {Systems {Biology} of {Pattern}
  {Formation}},\ \textbf {\bibinfo {volume} {460}},\ \bibinfo {pages} {2}
  (\bibinfo {year} {2020})}\BibitemShut {NoStop}%
\bibitem [{\citenamefont {Seshasai}\ \emph {et~al.}(2020)\citenamefont
  {Seshasai}, \citenamefont {Kurle}, \citenamefont {Verma},\ and\ \citenamefont
  {Pushpavanam}}]{seshasai_multiplicity_2020}%
  \BibitemOpen
  \bibfield  {author} {\bibinfo {author} {\bibfnamefont {P.~C.}\ \bibnamefont
  {Seshasai}}, \bibinfo {author} {\bibfnamefont {M.~M.}\ \bibnamefont {Kurle}},
  \bibinfo {author} {\bibfnamefont {N.}~\bibnamefont {Verma}}, \ and\ \bibinfo
  {author} {\bibfnamefont {S.}~\bibnamefont {Pushpavanam}},\ }\href {\doibase
  10.1016/j.ces.2020.115565} {\bibfield  {journal} {\bibinfo  {journal}
  {Chemical Engineering Science}\ }\textbf {\bibinfo {volume} {218}},\ \bibinfo
  {pages} {115565} (\bibinfo {year} {2020})}\BibitemShut {NoStop}%
\bibitem [{\citenamefont {Gray}\ and\ \citenamefont
  {Scott}(1983)}]{gray_autocatalytic_1983}%
  \BibitemOpen
  \bibfield  {author} {\bibinfo {author} {\bibfnamefont {P.}~\bibnamefont
  {Gray}}\ and\ \bibinfo {author} {\bibfnamefont {S.~K.}\ \bibnamefont
  {Scott}},\ }\href {\doibase 10.1016/0009-2509(83)80132-8} {\bibfield
  {journal} {\bibinfo  {journal} {Chemical Engineering Science}\ }\textbf
  {\bibinfo {volume} {38}},\ \bibinfo {pages} {29} (\bibinfo {year}
  {1983})}\BibitemShut {NoStop}%
\bibitem [{\citenamefont {Hethcote}(1989)}]{hethcote_three_1989}%
  \BibitemOpen
  \bibfield  {author} {\bibinfo {author} {\bibfnamefont {H.~W.}\ \bibnamefont
  {Hethcote}},\ }in\ \href {\doibase 10.1007/978-3-642-61317-3_5} {\emph
  {\bibinfo {booktitle} {Applied Mathematical Ecology}}},\ \bibinfo {editor}
  {edited by\ \bibinfo {editor} {\bibfnamefont {S.~A.}\ \bibnamefont {Levin}},
  \bibinfo {editor} {\bibfnamefont {T.~G.}\ \bibnamefont {Hallam}}, \ and\
  \bibinfo {editor} {\bibfnamefont {L.~J.}\ \bibnamefont {Gross}}}\ (\bibinfo
  {publisher} {Springer},\ \bibinfo {address} {Berlin, Heidelberg},\ \bibinfo
  {year} {1989})\ pp.\ \bibinfo {pages} {119--144}\BibitemShut {NoStop}%
\bibitem [{\citenamefont {Adamatzky}(2010)}]{adamatzky2010game}%
  \BibitemOpen
  \bibfield  {author} {\bibinfo {author} {\bibfnamefont {A.}~\bibnamefont
  {Adamatzky}},\ }\href@noop {} {\emph {\bibinfo {title} {Game of life cellular
  automata}}},\ Vol.~\bibinfo {volume} {1}\ (\bibinfo  {publisher} {Springer},\
  \bibinfo {year} {2010})\BibitemShut {NoStop}%
\bibitem [{\citenamefont {Zhang}\ and\ \citenamefont
  {Holden}(1995)}]{zhang_chaotic_1995}%
  \BibitemOpen
  \bibfield  {author} {\bibinfo {author} {\bibfnamefont {H.}~\bibnamefont
  {Zhang}}\ and\ \bibinfo {author} {\bibfnamefont {A.~V.}\ \bibnamefont
  {Holden}},\ }\href {\doibase 10.1016/0960-0779(93)E0048-G} {\bibfield
  {journal} {\bibinfo  {journal} {Chaos, Solitons \& Fractals}\ }\bibinfo
  {series} {Nonlinear {Phenomena} in {Excitable} {Physiological} {Systems}},\
  \textbf {\bibinfo {volume} {5}},\ \bibinfo {pages} {661} (\bibinfo {year}
  {1995})}\BibitemShut {NoStop}%
\bibitem [{\citenamefont {Coullet}\ \emph {et~al.}(1989)\citenamefont
  {Coullet}, \citenamefont {Gil},\ and\ \citenamefont
  {Lega}}]{coullet_defect-mediated_1989}%
  \BibitemOpen
  \bibfield  {author} {\bibinfo {author} {\bibfnamefont {P.}~\bibnamefont
  {Coullet}}, \bibinfo {author} {\bibfnamefont {L.}~\bibnamefont {Gil}}, \ and\
  \bibinfo {author} {\bibfnamefont {J.}~\bibnamefont {Lega}},\ }\href {\doibase
  10.1103/PhysRevLett.62.1619} {\bibfield  {journal} {\bibinfo  {journal}
  {Physical Review Letters}\ }\textbf {\bibinfo {volume} {62}},\ \bibinfo
  {pages} {1619} (\bibinfo {year} {1989})}\BibitemShut {NoStop}%
\bibitem [{\citenamefont {Epstein}\ and\ \citenamefont
  {Xu}(2016)}]{epstein2016reaction}%
  \BibitemOpen
  \bibfield  {author} {\bibinfo {author} {\bibfnamefont {I.~R.}\ \bibnamefont
  {Epstein}}\ and\ \bibinfo {author} {\bibfnamefont {B.}~\bibnamefont {Xu}},\
  }\href@noop {} {\bibfield  {journal} {\bibinfo  {journal} {Nature
  nanotechnology}\ }\textbf {\bibinfo {volume} {11}},\ \bibinfo {pages} {312}
  (\bibinfo {year} {2016})}\BibitemShut {NoStop}%
\bibitem [{\citenamefont {Greenberg}\ and\ \citenamefont
  {Hastings}(1978)}]{greenberg1978spatial}%
  \BibitemOpen
  \bibfield  {author} {\bibinfo {author} {\bibfnamefont {J.~M.}\ \bibnamefont
  {Greenberg}}\ and\ \bibinfo {author} {\bibfnamefont {S.~P.}\ \bibnamefont
  {Hastings}},\ }\href@noop {} {\bibfield  {journal} {\bibinfo  {journal} {SIAM
  Journal on Applied Mathematics}\ }\textbf {\bibinfo {volume} {34}},\ \bibinfo
  {pages} {515} (\bibinfo {year} {1978})}\BibitemShut {NoStop}%
\bibitem [{\citenamefont {Zakany}\ and\ \citenamefont
  {Milinkovitch}(2023)}]{zakany_simple_2023}%
  \BibitemOpen
  \bibfield  {author} {\bibinfo {author} {\bibfnamefont {S.}~\bibnamefont
  {Zakany}}\ and\ \bibinfo {author} {\bibfnamefont {M.~C.}\ \bibnamefont
  {Milinkovitch}},\ }\href {\doibase 10.1103/PhysRevX.13.041011} {\bibfield
  {journal} {\bibinfo  {journal} {Physical Review X}\ }\textbf {\bibinfo
  {volume} {13}},\ \bibinfo {pages} {041011} (\bibinfo {year}
  {2023})}\BibitemShut {NoStop}%
\bibitem [{\citenamefont {Boeynaems}\ \emph {et~al.}(2018)\citenamefont
  {Boeynaems}, \citenamefont {Alberti}, \citenamefont {Fawzi}, \citenamefont
  {Mittag}, \citenamefont {Polymenidou}, \citenamefont {Rousseau},
  \citenamefont {Schymkowitz}, \citenamefont {Shorter}, \citenamefont
  {Wolozin}, \citenamefont {Van Den~Bosch} \emph
  {et~al.}}]{boeynaems2018protein}%
  \BibitemOpen
  \bibfield  {author} {\bibinfo {author} {\bibfnamefont {S.}~\bibnamefont
  {Boeynaems}}, \bibinfo {author} {\bibfnamefont {S.}~\bibnamefont {Alberti}},
  \bibinfo {author} {\bibfnamefont {N.~L.}\ \bibnamefont {Fawzi}}, \bibinfo
  {author} {\bibfnamefont {T.}~\bibnamefont {Mittag}}, \bibinfo {author}
  {\bibfnamefont {M.}~\bibnamefont {Polymenidou}}, \bibinfo {author}
  {\bibfnamefont {F.}~\bibnamefont {Rousseau}}, \bibinfo {author}
  {\bibfnamefont {J.}~\bibnamefont {Schymkowitz}}, \bibinfo {author}
  {\bibfnamefont {J.}~\bibnamefont {Shorter}}, \bibinfo {author} {\bibfnamefont
  {B.}~\bibnamefont {Wolozin}}, \bibinfo {author} {\bibfnamefont
  {L.}~\bibnamefont {Van Den~Bosch}},  \emph {et~al.},\ }\href@noop {}
  {\bibfield  {journal} {\bibinfo  {journal} {Trends in cell biology}\ }\textbf
  {\bibinfo {volume} {28}},\ \bibinfo {pages} {420} (\bibinfo {year}
  {2018})}\BibitemShut {NoStop}%
\bibitem [{\citenamefont {Keating}(2012)}]{keating2012aqueous}%
  \BibitemOpen
  \bibfield  {author} {\bibinfo {author} {\bibfnamefont {C.~D.}\ \bibnamefont
  {Keating}},\ }\href@noop {} {\bibfield  {journal} {\bibinfo  {journal}
  {Accounts of chemical research}\ }\textbf {\bibinfo {volume} {45}},\ \bibinfo
  {pages} {2114} (\bibinfo {year} {2012})}\BibitemShut {NoStop}%
\end{thebibliography}%

\appendix
\clearpage
\onecolumngrid
\setcounter{figure}{0} % Reset figure counter
\renewcommand{\thefigure}{S\arabic{figure}} % Prefix figure numbers with "S"
\renewcommand{\thesection}{S\arabic{section}} % Prefix figure numbers with "S"

\section{Nondimensionalization of the models}
\subsection{Model (M1)}
We begin with Model (M1), in which only species $P$ phase separates, and species $Z$ exhibits diffusive dynamics.

Beginning from the kinetic equations:
\begin{align} \label{eq:fund}
    &\frac{\partial c_P(\mathbf x,t)}{\partial t}=D_P\nabla^2 \left(\frac{\delta F(c_P,c_Z)}{\delta c_P}\right)+R_P(c_P,c_Z) \\
    &\frac{\partial c_Z(\mathbf x,t)}{\partial t}=D_Z\nabla^2 \left(\frac{\delta F(c_P,c_Z)}{\delta c_Z}\right)+R_Z(c_P,c_Z)
\end{align}

Inserting the model (M1) free energy:
\begin{align} \label{eq:fund}
    &\frac{\partial c_P(\mathbf x,t)}{\partial t}=D_P\nabla^2 \left(a_1 c_P(\mathbf x,t)+a_2 c_P(\mathbf x,t)^2+a_3 c_P(\mathbf x,t)^3+\epsilon c_Z(\mathbf x,t)\right)-D_P \gamma^2 \nabla^4 c_P + R_P(c_P(\mathbf x,t),c_Z(\mathbf x,t)) \\
    &\frac{\partial c_Z(\mathbf x,t)}{\partial t}=D_Z\nabla^2 \left(c_Z(\mathbf x,t) +\epsilon c_P(\mathbf x,t)\right)+R_Z(c_P(\mathbf x,t),c_Z(\mathbf x,t))
\end{align}

inserting non-dimensional density, space and time, we have:
\begin{align}
c_P&=c_P^{(0)} \tilde c_P \\
c_Z&=c_Z^{(0)} \tilde c_Z \\
x &= x_c \xi \\
t &= t_c \tau
\end{align}
leading to (suppressing dependence on space and time):
\begin{align} \label{eq:fund}
    &\frac{c_P^{(0)}}{t_c}\frac{\partial \tilde c_P}{\partial \tau}=\frac{ D_P c_P^{(0)}}{x_c^2}\nabla^2 \left(a_1 \tilde c_P+c_P^{(0)} a_2 \tilde c_P^2+a_3(c_P^{(0)})^2 \tilde c_P^3+\epsilon \frac{c_Z^{(0)}}{c_P^{(0)}}\tilde c_Z\right)-\frac{D_P c_P^{(0)}\gamma^2}{x_c^4} \nabla^4 \tilde c_P + R_P(\tilde c_P,\tilde c_Z) \\
    &\frac{c_Z^{(0)}}{t_c}\frac{\partial \tilde c_Z}{\partial \tau}=\frac{D_Z c_Z^{(0)}}{x_c^2}\nabla^2 \left(\tilde c_Z +\epsilon \frac{c_P^{(0)}}{c_Z^{(0)}} \tilde c_P\right)+R_Z(\tilde c_P,\tilde c_Z)
\end{align}
where now $\nabla=\frac{\partial}{\partial \xi}$.
now we set these parameters as follows:
\begin{align}
x_c &=\gamma \\
t_c &=\gamma^2/D_P \\
\end{align}
and dividing through everything by the prefactors in front of the time derivatives, we have:
\begin{align} \label{eq:fund}
    &\frac{\partial \tilde c_P}{\partial \tau}=\nabla^2 \left(a_1 \tilde c_P+c_P^{(0)} a_2 \tilde c_P^2+a_3(c_P^{(0)})^2 \tilde c_P^3+\epsilon \frac{c_Z^{(0)}}{c_P^{(0)}}\tilde c_Z\right)-\nabla^4 \tilde c_P + \tilde R_P(\tilde c_P,\tilde c_Z) \\
    &\frac{\partial \tilde c_Z}{\partial \tau}=\frac{D_Z}{D_P}\nabla^2 \left(\tilde c_Z +\epsilon \frac{c_P^{(0)}}{c_Z^{(0)}} \tilde c_P\right)+\tilde R_Z(\tilde c_P,\tilde c_Z)
\end{align}

and then defining the parameters
\begin{align}
\delta &= D_Z/D_P \\
\rho &= c_P^{(0)}/c_Z^{(0)} \\
\tilde a_2 & = c_P^{(0)} a_2 \\
\tilde a_3 &= (c_P^{(0)})^2 a_3 \\
\tilde R_P&=\frac{\gamma^2}{D_P c_P^{(0)}}R_P(c_P^{(0)} \tilde c_P,c_Z^{(0)} \tilde c_Z) \\
\tilde R_Z&=\frac{\gamma^2}{D_P c_Z^{(0)}}R_Z(c_P^{(0)} \tilde c_P,c_Z^{(0)} \tilde c_Z)
\end{align}
we have reduced the number of parameters of the problem.

\begin{align} \label{eq:fund}
    &\frac{\partial \tilde c_P}{\partial \tau}=\nabla^2 \left(a_1 \tilde c_P+\tilde a_2 \tilde c_P^2+\tilde a_3 \tilde c_P^3+(\epsilon/\rho) \tilde c_Z\right)-\nabla^4 \tilde c_P + \tilde R_P(\tilde c_P,\tilde c_Z) \\
    &\frac{\partial \tilde c_Z}{\partial \tau}=\delta \nabla^2 \left(\tilde c_Z +\epsilon \rho \tilde c_P\right)+\tilde R_Z(\tilde c_P,\tilde c_Z)
\end{align}

\subsection{Model (M2)} 
Beginning from the equations:
\begin{align} \label{eq:fund}
    &\frac{\partial c_P(\mathbf x,t)}{\partial t}=D_P\nabla^2 \left(a_1 c_P(\mathbf x,t)+a_2 c_P(\mathbf x,t)^2+a_3 c_P(\mathbf x,t)^3+\epsilon c_Z(\mathbf x,t)\right)-D_P \gamma_P^2 \nabla^4 c_P + R_P(c_P(\mathbf x,t),c_Z(\mathbf x,t)) \\
    &\frac{\partial c_Z(\mathbf x,t)}{\partial t}=D_Z\nabla^2 \left(a_1 c_Z(\mathbf x,t)+b_2 c_Z(\mathbf x,t)^2+b_3 c_Z(\mathbf x,t)^3+\epsilon c_P(\mathbf x,t)\right)-D_Z \gamma_Z^2 \nabla^4 c_Z+R_Z(c_P(\mathbf x,t),c_Z(\mathbf x,t))
\end{align}

Using the same nondimensional units defined in the previous section, and introducing the notation $\Pi_P(\tilde c_P)=\frac{1}{c_P^{0}}\left(a_1 c_P(\mathbf x,t)+a_2 c_P(\mathbf x,t)^2+a_3 c_P(\mathbf x,t)^3\right),\Pi_Z(\tilde c_P)=\frac{1}{c_Z^{0}}\left(b_1 c_Z(\mathbf x,t)+b_2 c_Z(\mathbf x,t)^2+b_3 c_Z(\mathbf x,t)^3\right)$ we have the following equations:

\begin{align} \label{eq:fund}
    &\frac{\partial \tilde c_P}{\partial \tau}=\nabla^2 \left(\Pi_P(\tilde c_P)+(\epsilon/\rho) \tilde c_Z\right)-\nabla^4 \tilde c_P + \tilde R_P(\tilde c_P,\tilde c_Z) \\
    &\frac{\partial \tilde c_Z}{\partial \tau}=\delta \nabla^2 \left(\Pi_Z(\tilde c_Z)+\epsilon \rho \tilde c_P(\mathbf x,t)\right)-\delta \frac{\gamma_Z^2}{\gamma_P^2} \nabla^4 \tilde c_Z+\tilde R_Z(\tilde c_P,\tilde c_Z)
\end{align}

\section{Conditions for appearance of imaginary eigenvalues}

In order to calculate the conditions necessary for the appearance of imaginary eigenvalues, we can utilize all the conditions shown in the main text, which we repeat here:
\begin{align}
\boldsymbol{v^T} .\underline \Omega =\text{det}(\underline \Omega)&=0 \text{ (mass conservation) } \label{eq:cond1}\\
\text{Tr}(\underline  \Omega)&<0 \,\text{ (stable chemical fixed point)} \label{eq:cond2}\\
\text{Tr}(\underline \Delta(\mathbf k) + \underline \Omega) &>0 \text{ (growth instability) } \label{eq:cond3}\\
\text{det}(\underline \Delta(\mathbf k) + \underline \Omega) &>\frac{1}{4}\text{Tr}(\underline \Delta(\mathbf k) + \underline \Omega)^2 \text{ (oscillations) } \label{eq:cond4}
\end{align}

It can be seen that combination of conditions \eqref{eq:cond1}-\eqref{eq:cond3} imply that for some finite $k$, the sum of the two diagonal components:
\begin{align}
a_{PP}(\mathbf k)=& -k^2\left( \frac{\mathrm d \Pi_P(\tilde c_P)}{\mathrm d \tilde c_P}\bigg|_{\tilde c_P*}+k^2\right)\\
a_{ZZ}(\mathbf k)=& -k^2\left( \frac{\mathrm d \Pi_Z(\tilde c_Z)}{\mathrm d \tilde c_Z}\bigg|_{\tilde c_Z*}+ g^2 k^2\right)
\end{align}
must be positive, this implies that:
\begin{equation}
     \frac{\mathrm d \Pi_P(\tilde c_P)}{\mathrm d \tilde c_P}\bigg|_{\tilde c_P*}+\frac{\mathrm d \Pi_Z(\tilde c_Z)}{\mathrm d \tilde c_Z}\bigg|_{\tilde c_Z*} < -k^2 - g^2 k^2
\end{equation}
As $g$ and $k$ are positive, this means that the two terms on the left hand side must be negative. We take $k=1$ which is the same as choosing a length scale in our system. We call $\frac{\mathrm d \Pi_P(\tilde c_P)}{\mathrm d \tilde c_P}\bigg|_{\tilde c_P*}=p_P$ and $\frac{\mathrm d \Pi_Z(\tilde c_Z)}{\mathrm d \tilde c_Z}\bigg|_{\tilde c_Z*}=p_b$ . Rule \eqref{eq:cond4} can be put in component form as:
\begin{equation} \label{eq:cond5}
4 \delta  \chi ^2+2 J_{11} \left(\delta  g^2+J_{12}+p_P-\delta  p_b+2 \chi
   -1\right)+\left(-\delta  g^2+J_{12}-p_P+\delta  p_b+1\right){}^2+4 \delta
    J_{12} \chi +J_{11}^2<0
\end{equation}
where the components $J$ arise from the matrix $\Omega$:
\begin{equation}
\Omega = \begin{bmatrix}
J_{11} & -J_{12} \\
-J_{11} & J_{12} \\
\end{bmatrix}
\end{equation}
where the structure of the matrix arises from the fact that it's a closed system with a conservation law. Another consequence of this is that $J_{11}+J_{12}<0$. Equation \eqref{eq:cond5} is simple enough to state but not to visualize, however it can be seen that for negative values of $J_{11}$ and $J_{12}$ (as is normally the case), that it is better for $\chi$ to be positive.

The structure of chemical reactions for achieving imaginary eigenvalues can be more easily discerned by rewriting expression equation \eqref{eq:cond5} as:
\begin{equation}
a J_{11}+b J_{12}+c+J_{11}^2+2 J_{12} J_{11}+J_{12}^2<0    
\end{equation}

where $a,b,c$ are constants derived from \eqref{eq:cond5}, and it can be shown that $c>0$. It can be seen when $J_{11}$ and $J_{12}$ are opposite signs that this inequality is more likely to be fulfilled, depending on the values of $a$ and $b$. Furthermore, $a,b$ have a special dependence on one another given by $a=d-4 \chi$ and $b=-d+4\chi \delta$ where d is another constant. It is not possible for $J_{11}$ and $J_{12}$ to take opposite signs in a closed system with mass conservation. However, an autocatalytic cycle can have one of $J_{11}$ or $J_{12}$ equal to zero.

% \subsection{Radial equations}

% The above models can also be recast in radial coordinates $(\chi,\psi)\to(r,\theta)$. For simplicity, we assume solutions with azimuthal symmetry. In other the words, the dimensionless concentration $\tilde c_P(\chi,\psi,\tau)=\tilde c_P(r,\theta,\tau)=\tilde c_P(r,\tau)$ (where $r$ is also dimensionless)

% This leads to the following transformations of the operators involved in the equations:
% \begin{align}
%     \nabla^2 &=\frac{\partial^2}{\partial r^2}+\frac{1}{r}\frac{\partial}{\partial r} \\
%     \nabla^4 &=\frac{\partial^4}{\partial r^4}+\frac{2}{r}\frac{\partial^3}{\partial r^3}-\frac{1}{r^2}\frac{\partial^2}{\partial r^2}+\frac{1}{r^3}\frac{\partial}{\partial r}
% \end{align}

\section{Analysis of conditions for appearance of dynamics}

In the main text, we had conditions for the appearance of growing dynamically oscillating states, given by:

\begin{align}
\boldsymbol{v^T} .\underline J =\text{det}(\underline J)&=0 \text{ (mass conservation) } \label{eq:cond1}\\
\text{Tr}(\underline J)&<0 \,\text{ (stable chemical fixed point)} \label{eq:cond2}\\
\text{Tr}(\underline \Delta(\mathbf k) + \underline J) &>0 \text{ (growth instability) } \label{eq:cond3}\\
\text{det}(\underline \Delta(\mathbf k) + \underline J) &>\frac{1}{4}\text{Tr}(\underline \Delta(\mathbf k) + \underline J)^2 \text{ (oscillations) } \label{eq:cond4}
\end{align}
condition \eqref{eq:cond1} translates to a matrix which must be given by:
\begin{equation}
\underline J = \begin{bmatrix}
J_{PP} & -J_{PZ} \\
-J_{PP} & J_{PZ} \\
\end{bmatrix}
\end{equation}
where $J_{PP}\le0,J_{PZ}\le0$ in a closed system with mass action kinetics.

where in component form the matrix $M=\underline \Delta(\mathbf k) + \underline J$ is given by:

\begin{equation}
\underline M = \left(
\begin{array}{cc}
 J_{PP}-\mathbf k^2 \left(\mathbf k^2-p_a\right) & -\frac{J_{PZ}}{\rho }-\frac{\mathbf k^2 \chi }{\rho } \\
 -J_{PP} \rho -\delta  \mathbf k^2 \rho  \chi  & J_{PZ}-\delta  \mathbf k^2 \left(g ^2 \mathbf k^2-p_b\right) \\
\end{array}
\right)
\end{equation}
where we have used the fact that for a two chemical component system where the concentrations are conserved to write $J_{ZP}=-J_{PP}$ and $J_{ZZ}=J_{PZ}$. We have also introduced the short form that 
\begin{align}
p_a &= \frac{\mathrm d \Pi_P(\tilde c_P)}{\mathrm d \tilde c_P}\bigg|_{\tilde c_P*}\\
p_b &=\frac{\mathrm d \Pi_Z(\tilde c_Z)}{\mathrm d \tilde c_Z}\bigg|_{\tilde c_Z*}
\end{align}

One can then analyze the conditions on the matrices in terms of all the parameters appearing in the above matrix, and find the regions in parameter space where dynamics occurs. This is easiest to visualize by plotting the conditions as a function which equals 1 when all the conditions are true, and zero when all the conditions are false. Which we do below:

\begin{figure}[!h]
    \includegraphics[width=90mm]{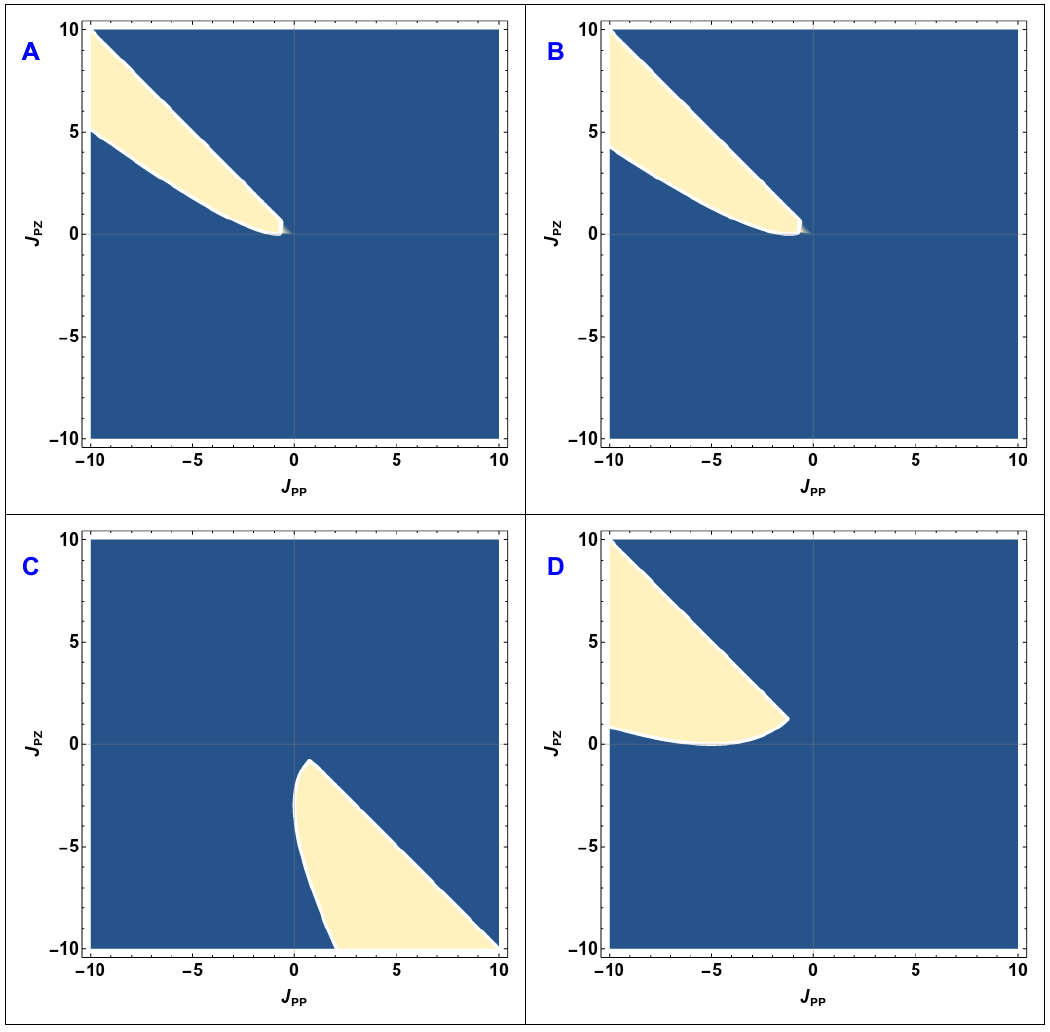}
    \caption{Regions where all conditions \eqref{eq:cond1}-\eqref{eq:cond4} are true for  $p_a=-2$,$p_b=-2$,$|\mathbf k|=1$,$\chi=0$ and A) $\delta =0.1, g=0.1$,B) $\delta =0.1, g=2$,C) $\delta =2, g=0.1$,D) $\delta =2, g=2$ }
    \label{fig:figg1}
\end{figure}

\begin{figure}[!h]
    \includegraphics[width=90mm]{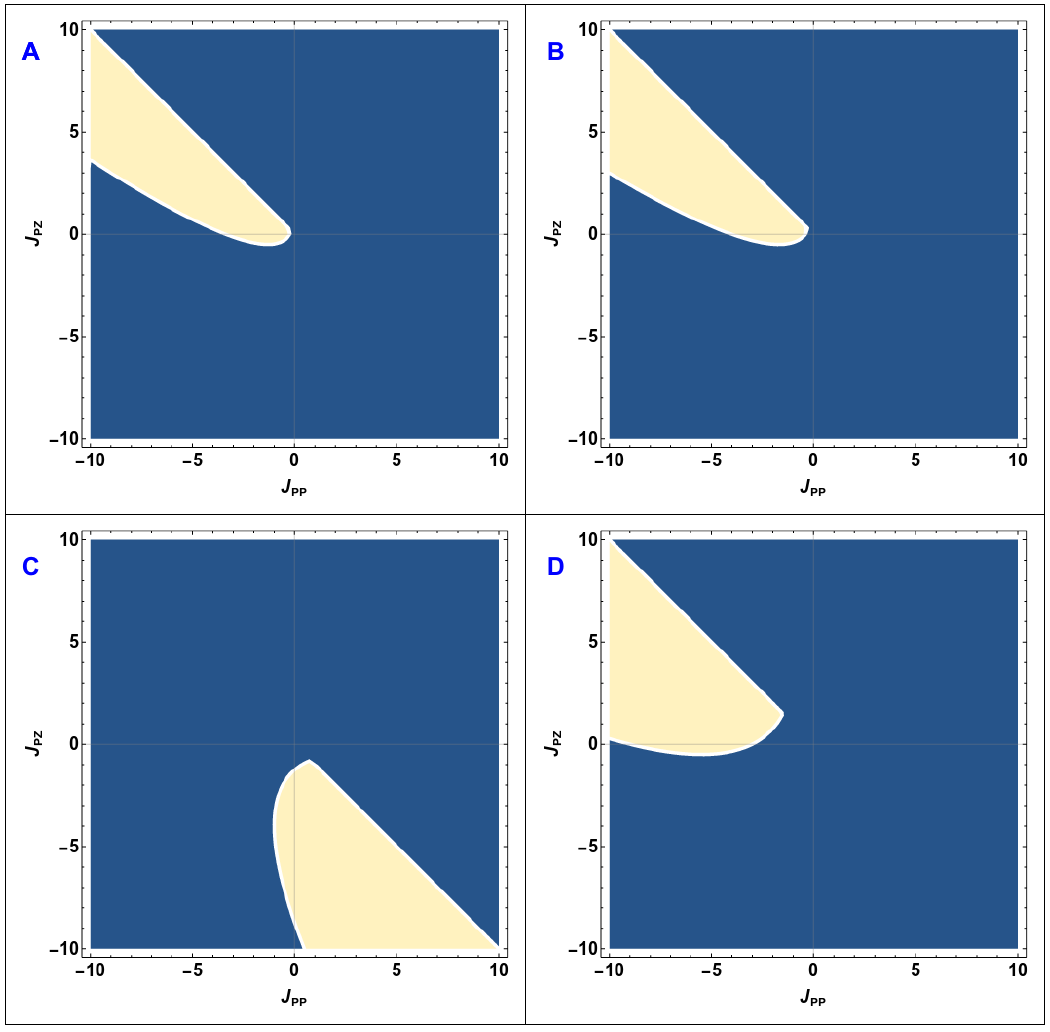}
    \caption{Regions where all conditions \eqref{eq:cond1}-\eqref{eq:cond4} are true for  $p_a=-2$,$p_b=-2$,$|\mathbf k|=1$,$\chi=0.5$ and A) $\delta =0.1, g=0.1$,B) $\delta =0.1, g=2$,C) $\delta =2, g=0.1$,D) $\delta =2, g=2$}
    \label{fig:figg2}
\end{figure}

\begin{figure}[!h]
    \includegraphics[width=90mm]{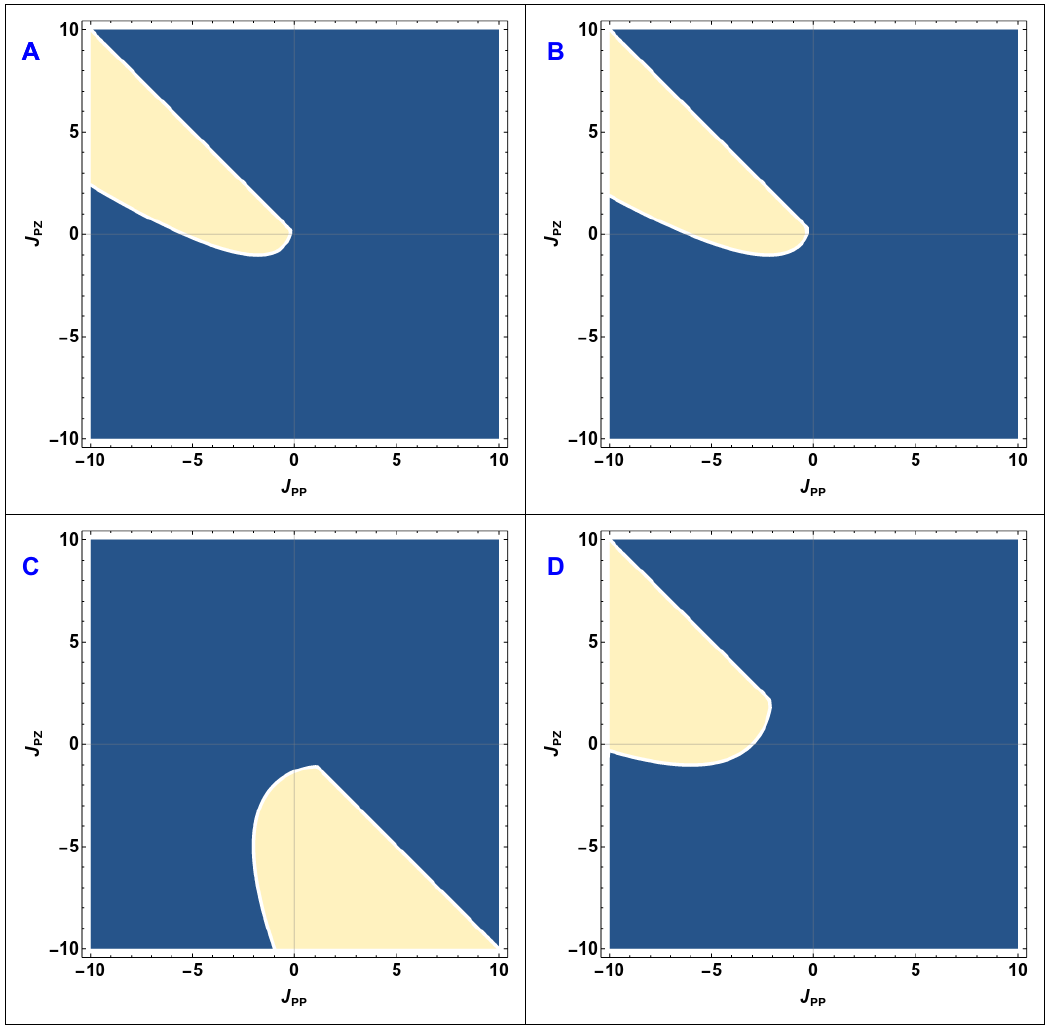}
    \caption{Regions where all conditions \eqref{eq:cond1}-\eqref{eq:cond4} are true for  $p_a=-2$,$p_b=-2$,$|\mathbf k|=1$,$\chi=1$ and A) $\delta =0.1, g=0.1$,B) $\delta =0.1, g=2$,C) $\delta =2, g=0.1$,D) $\delta =2, g=2$}
    \label{fig:figg3}
\end{figure}

\clearpage

In the main text, it was said that autocatalytic reactions can more robustly produce dynamical phenomena in our system. We demonstrate this 

\section{Numerical Details}

When numerically solving the PDEs in the main text, we use a Fourier spectral method. We choose Von-Neumann boundary conditions $\partial_x c(x,t)=0, x\to L$, and the initial state is given by a random perturbation of the chemically stable homogeneous state, with variation given uniformly up to 50 to 150\% of the fixed point.

For a generic equation of the form:

\begin{equation}
\frac{\partial c(\mathbf x,t)}{\partial t} =  \nabla^2 (f(c(\mathbf x,t)))+\nabla^4 (Q c(\mathbf x,t))+R(c(\mathbf x,t))
\end{equation}

we use the following discretization:

\begin{equation}
\frac{\mathbf c^{n+1}-\mathbf c^{n} }{\Delta t} = \nabla^2 (f(\mathbf c^n))+ \nabla^4 (Q \mathbf c^{n+1})+R(\mathbf c^n)
\end{equation}

where the vector $\mathbf c^n$ refers to the collection of all the points in space at a time $n \Delta t$.

In the absence of a $\mathcal{O}(\nabla^4)$ term, the linear part of the term in $\nabla ^2$ is taken to be at $n+1$ instead. These sets of equations can be solved via standard linear solvers in Fourier space, by first calculating $f(\mathbf c^n)$ and $R(\mathbf c^n)$ and then taking their Fourier transforms and using the well known properties of Fourier transforms of derivatives. Different discretization schemes did not affect our results. We also modified the size of the mesh and changed the timestep $\Delta t$ over many different orders of magnitude as a numerical check on our observations. Simulations were allowed to run for $>10^6$ timesteps until the solution appeared settled into a dynamical state. We also verified our identification of different phases by starting with different initial conditions (such as starting with a spiral in the case where the disordered droplets occured).

The results in the main text were presented with Von-Neumann boundary conditions. Simulations were also performed for periodic boundary conditions, though the qualitative results remained the same.

\clearpage 
\section{Additional figures for models (M1) and (M2)}

\begin{figure}[!h]
    \centering
    \includegraphics{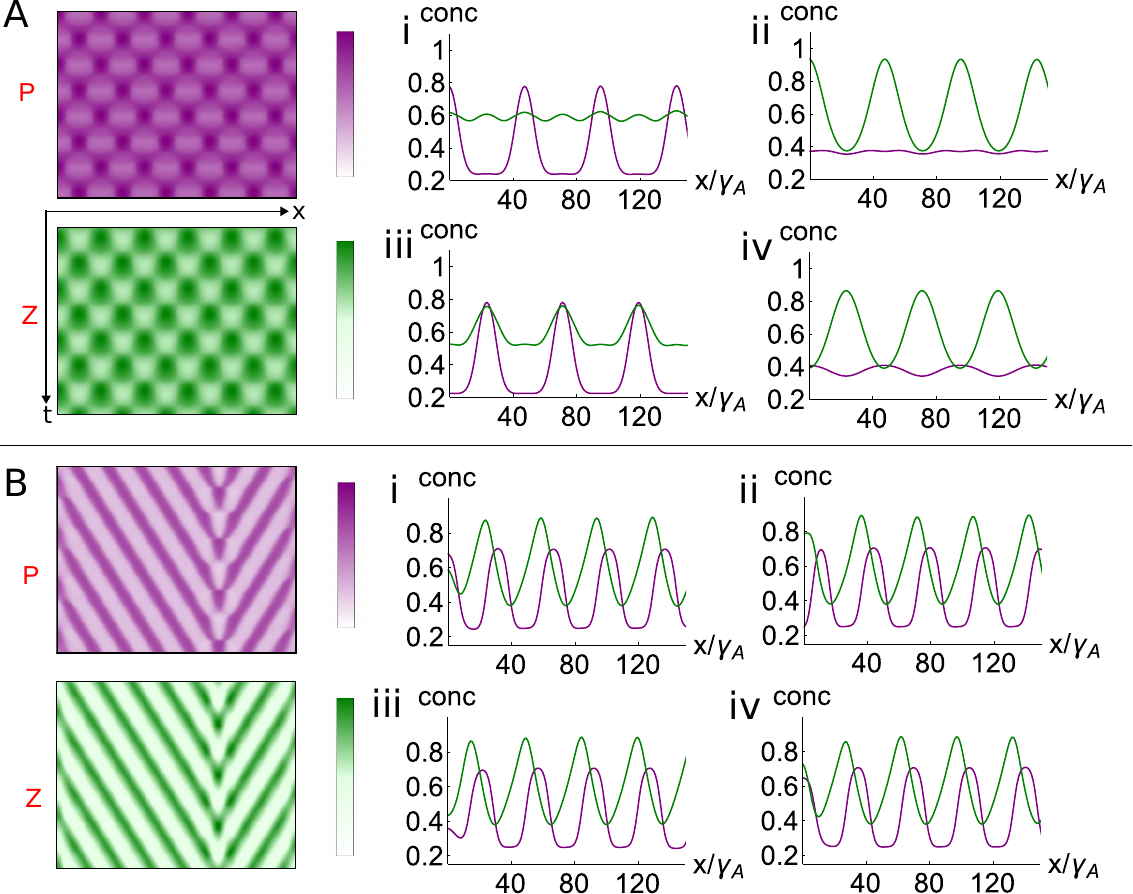}
    \caption{Dynamical behaviors observed in model (M1) in one dimension (to be compared to Fig. 1 in the main text). (A) Standing waves (pulses) are observed, with slightly different dynamics than model (M2), (B) Travelling wave trains can also be observed in model (M1). }
    \label{fig:M1D}
\end{figure}

\begin{figure}[!h]
    \includegraphics[width=120mm]{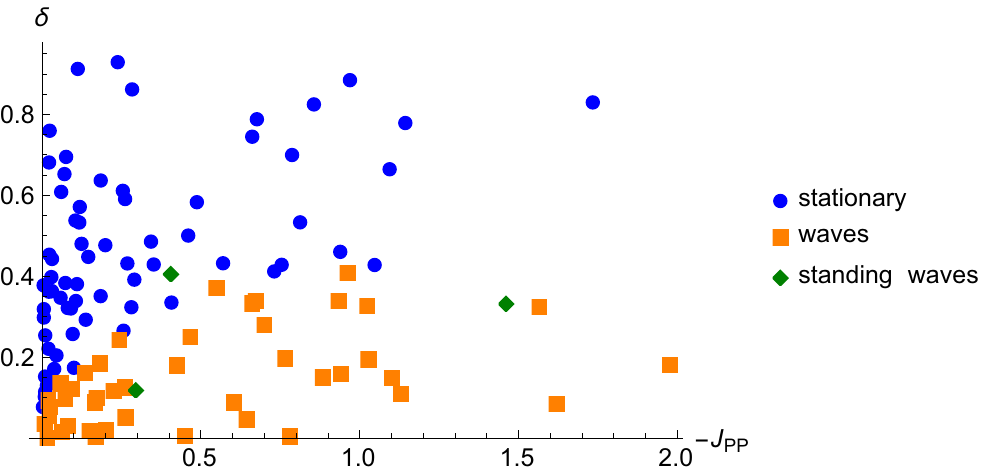}
    \caption{Phase diagram of observed dynamical behaviors as a function of the effective Jacobian parameter $J_{PP}$ arising from chemistry and the ratio of the diffusion coefficients $\delta$ for model (M1) where one species is not phase separating. This is a two dimensional slice where all the other parameters were randomly chosen, nevertheless one can observe a transition between a stationary state and travelling wave solutions as $\delta$ and  $J_{PP}$ are changed. Along the boundary, a few points are standing pulses. Note that as $g=0$ for model (M1), we instead change $J_{PP}$ and $\delta$  }
    \label{fig:M1PD}
\end{figure}

\begin{figure}[!h]
    \includegraphics[width=120mm]{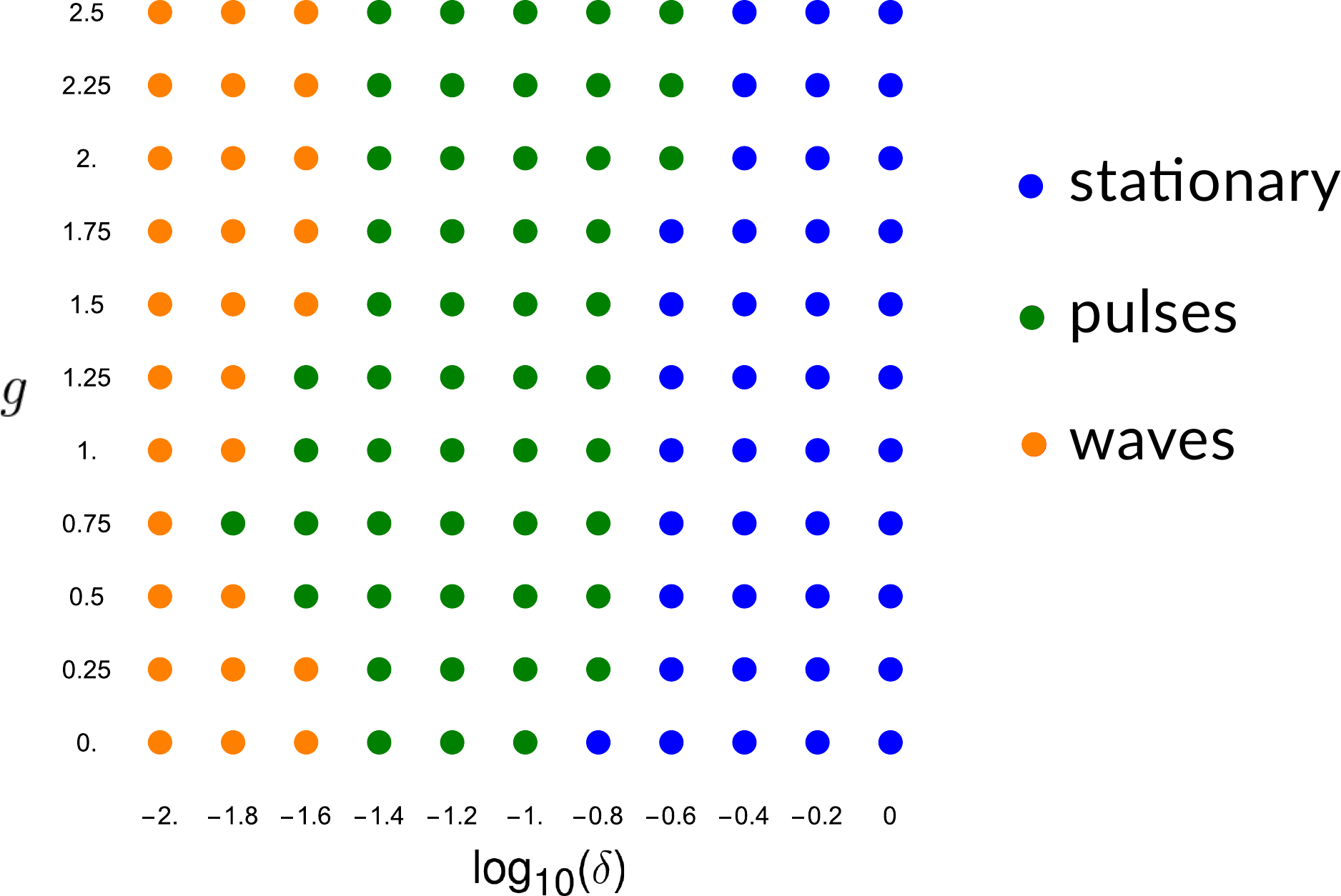}
    \caption{Phase diagram of observed dynamical behaviors as a function of the ratio of surface tensions and ratio of diffusion coefficients for model (M2). While the surface tensions have little effect on the dynamics, changing the diffusion constants leads to transitions from wave like behavior to the pulses described in the main text to situations where no dynamics occurs (stationary)}
    \label{fig:M2PD}
\end{figure}

\begin{figure}[!h]
    \includegraphics[width=120mm]{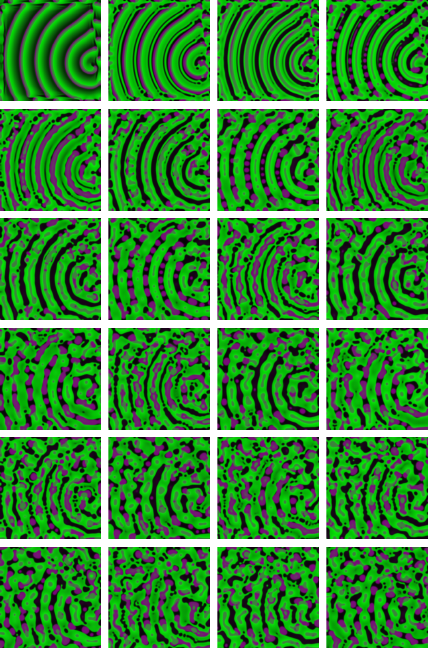}
    \caption{Time evolution of breakup of spiral wave when started in the parameters corresponding to disordered droplets. Time advances left to right, top to bottom. }
    \label{fig:M2SWB}
\end{figure}

\clearpage

\clearpage

\section{Table of Parameters Used in Figures 1-4 of the Manuscript}

For simplicity, when both species are phase separating (Figures 1, 2 and 3), we keep their phase separation parameters ($\rho_0,\rho_1,\nu$) the same. When one species does not phase separate (Figure 4) we list the phase separation parameters for the remaining phase separating species. 

\begin{table}[htbp]
\begin{tabular}{ |p{2cm}||p{1cm}|p{1cm}|p{1cm}|p{1.3cm}|p{1cm}|p{1cm}|p{1cm}|p{1cm}|p{1cm}|p{1cm}|  }
 \hline
 %\multicolumn{11}{|c|}{Parameter Table} \\
 \hline
 Figure & $\delta$ & $g$ & $L$ & $\Delta t$ & $k_1$ & $k_2$ & $\rho_0$ & $\rho_1$ & $\nu$ & $\chi$ \\
 \hline
 Fig 1A  & 0.01    &0.8&   40 & 0.00045 & 1.57 & 4.47 & 0.2 & 0.8 & 10 & 0.5\\ \hline
 Fig 1B  & 0.04    &0.8&   40 & 0.00045 & 1.57 & 4.47 & 0.2 & 0.8 & 10 & 0.5 \\ \hline
 Fig 1C  & 0.1    &0.8&   40 & 0.00045 & 1.57 & 4.47 & 0.2 & 0.8 & 10 & 0.5 \\ \hline
 Fig 2A & 0.01    &0.8&   40 & 0.00045 & 1.57 & 4.47 & 0.2 & 0.8 & 10 & 0.5\\ \hline
 Fig 2B & 0.04    &0.8&   40 & 0.00045 & 1.57 & 4.47 & 0.2 & 0.8 & 10 & 0.5 \\ \hline
 Fig 3& 0.1    &0.8&   40 & 0.00045 & 1.57 & 4.47 & 0.2 & 0.8 & 10 & 0.5 \\ \hline
 Fig 4& 0.34    &0.&   40 & 0.00045 & 4.32 & 9.47 & 0.23 & 0.936 & 9.19 & 0.5 \\ \hline
\end{tabular}
\caption{Table of parameters used in the simulations reported in the manuscript.}
\end{table}

\end{document}